\documentclass[12pt,a4paper]{article}
\usepackage{jcappub}
\usepackage[T1]{fontenc}
\usepackage{lmodern}
\usepackage{amsmath,amssymb,mathtools,bm,slashed}
\usepackage[version=4]{mhchem}
\usepackage{graphicx,booktabs,array,microtype,placeins,etoolbox}
\usepackage{orcidlink,xcolor}
\makeatletter
\patchcmd{\l@section}{1.0em}{0.75em}{}{\PackageError{manuscript}{Contents spacing patch failed}{Check the section-entry definition.}}
\makeatother
\let\fulltableofcontents\tableofcontents
\renewcommand{\tableofcontents}{\begingroup\small\setlength{\baselineskip}{11pt}\enlargethispage{2\baselineskip}\fulltableofcontents\endgroup}
\newcommand{\GeV}{\,\mathrm{GeV}}
\newcommand{\MeV}{\,\mathrm{MeV}}
\newcommand{\Mpl}{M_{\mathrm{Pl}}}
\newcommand{\dd}{\mathrm d}
\newcommand{\Veff}{V_{\mathrm{eff}}}
\newcommand{\TD}{T_D}

\newcommand{\bfrac}[2]{\left(\frac{#1}{#2} \right)  }

\graphicspath{{figures/}}

\title{Nanohertz gravitational waves and asymmetric dark matter from a supercooled dark sector phase transition}

\author[a,b]{Wonsub Cho\,\orcidlink{0000-0003-2493-4628}}
\emailAdd{sub526@skku.edu}
\author[a]{Ki-Young Choi\,\orcidlink{0000-0001-7604-6644}}
\emailAdd{kiyoungchoi@skku.edu}
\affiliation[a]{Department of Physics and Institute of Basic Science, Sungkyunkwan University, Suwon 16419, Korea}
\affiliation[b]{Theoretical Physics Department, CERN, 1 Esplanade des Particules, CH-1211 Geneva 23, Switzerland}
\abstract{We study the connection between asymmetric dark matter and nanohertz gravitational waves from a supercooled first-order phase transition in a classically conformal dark sector.
The dark sector contains a real scalar, a complex scalar and a Dirac fermion, whose mass arises from radiative symmetry breaking.
An additional Affleck--Dine condensate generates a fermion asymmetry and heats the dark bath before nucleation.
The conserved dark charge fixes the asymmetry between fermions and antifermions, while entropy production during delayed visible-sector reheating dilutes the relic and determines the redshift of the gravitational-wave spectrum.
We calculate the thermal nucleation rate in a three-dimensional effective theory with finite-density corrections, next-to-leading-order contributions and fluctuation determinants (NLOdet).
We follow the coupled thermal evolution of the dark and visible sectors.
For a benchmark chosen to match the observed dark-matter abundance, the transition completes at MeV-scale temperatures and produces a nanohertz signal under the adopted dynamical and thermal assumptions.
We compare the predicted collision-dominated gravitational-wave spectra with free-spectrum posteriors from pulsar timing arrays.
For a benchmark point and variations of the dark-sector parameters, the gravitational-wave spectra and dark matter relic abundances are compared, and compatibility with Big Bang nucleosynthesis is assessed.}
\begin{document}
\maketitle

\section{Introduction}
\label{sec:introduction}
Pulsar timing arrays (PTAs) search for a stochastic gravitational-wave background (SGWB) through timing residuals, the differences between observed and predicted arrival times of millisecond-pulsar signals.
An isotropic SGWB correlates the timing residuals of pulsar pairs with the characteristic Hellings--Downs dependence on their angular separation~\cite{Hellings:1983fr}.
In 2023, NANOGrav and EPTA--InPTA reported evidence for these correlations at nanohertz frequencies, with significance depending on the statistic~\cite{NANOGrav:2023gor,EPTA:2023fyk}.
For the $10.3$-year modern-system EPTA subset, the false-alarm probability was approximately $0.1\%$, corresponding to a significance of $\gtrsim3\sigma$.
Adding InPTA data gave broadly consistent results~\cite{EPTA:2023fyk}.
Complementary searches used PPTA DR3~\cite{Reardon:2023gzh} and CPTA DR1, the latter probing quadrupolar correlations with FAST observations~\cite{Xu:2023wog}.

The origin of the nanohertz background remains uncertain.
Inspiralling supermassive black hole binaries (SMBHBs) provide a conventional astrophysical interpretation~\cite{NANOGrav:2023hfp}, while possible cosmological sources include inflation, cosmic strings, domain walls and first-order phase transitions (FOPTs)~\cite{NANOGrav:2023hvm}.
A frequentist comparison of stochastic templates found a $2\text{--}3\sigma$ preference for FOPT spectra over two SMBHB population models~\cite{Winkler:2024olr}, but this depends on the population models used and is not a general conclusion of the PTA collaborations.
Discrete SMBHB populations can also produce departures from a single power law, providing another source of spectral structure~\cite{Agazie:2024jbf}.

The PTA signal is compatible with FOPTs at MeV to sub-GeV temperatures, depending on whether bubble collisions or sound waves dominate~\cite{NANOGrav:2023hvm}.
The QCD~\cite{Aoki:2006we} and Standard Model electroweak transitions~\cite{DOnofrio:2015gop} are smooth crossovers, motivating FOPTs in dark sectors weakly coupled to the visible sector~\cite{Schwaller:2015tja,Borah:2021ocu,Borah:2021ftr,Bringmann:2023opz,Fujikura:2023lkn,Fujikura:2024jto,Goncalves:2025uwh,Costa:2025csj}.
Nearly conformal dynamics can lead to strong supercooling and a period of vacuum domination, followed by reheating that changes particle abundances~\cite{Konstandin:2011dr}.
In a secluded supercooled sector, the percolation and reheating temperatures can differ, making the thermal history essential to predicting the gravitational-wave (GW) signal~\cite{Bringmann:2023opz}.

Previous studies have connected conformal phase transitions to both gravitational waves and dark matter.
In a classically conformal $B-L$ model, strong supercooling can enhance the GW signal~\cite{Jinno:2016knw}, although transition completion must be checked separately~\cite{Ellis:2018mja}.
Supercooling can also alter the dark-matter abundance through dilution and post-transition production~\cite{Hambye:2018qjv}.
A conformal gauge-sector model has been used to study fermion relics, reheating and PTA inference together~\cite{Balan:2025uke}.
A separate study used a gauge-independent effective action to describe nucleation in a cogenesis model~\cite{Feng:2025wvc}.
In another scenario, scalar oscillations after the transition produce particles whose decays generate a dark asymmetry~\cite{Girmohanta:2025wcq}.

Here we study how a dark asymmetry established before nucleation and the subsequent thermal evolution connect the relic abundance to the GW signal.
The dark sector contains a real scalar, a complex scalar $S$ and a Dirac fermion, with a classically conformal scalar potential and a Yukawa interaction.
Loops of $S$ generate symmetry breaking radiatively~\cite{Coleman:1973jx,Gildener:1976ih}, giving the fermion its mass and allowing a supercooled transition.
The same scalar potential and Yukawa interaction were previously used to study scalar dark-matter freeze-in after supercooling, in a model with additional Higgs and lepton portals~\cite{Wong:2023qon}.

We impose equal initial temperatures for the dark and visible baths.
An additional Affleck--Dine (AD) condensate transfers charge and energy to the dark bath before nucleation~\cite{Affleck:1984fy}.
Together with the primordial bath, this transfer sets the temperature and fermion chemical potential in the false-vacuum phase.
After the transition, pair annihilation preserves the asymmetry between fermions and antifermions~\cite{Kaplan:2009ag,Iminniyaz:2011yp}.
Entropy production during delayed visible-sector reheating dilutes this asymmetric relic and determines the GW redshift.
The relic abundance and observable GW spectrum are therefore linked through the conserved dark charge and the same reheating history.

We compute the thermal nucleation rate in a three-dimensional effective theory~\cite{Puchades-Ibanez:2026ksh,Kierkla:2026bnm} and follow the coupled evolution through the transition and visible-sector reheating.
A benchmark chosen to match the observed dark-matter abundance completes the transition and produces a nanohertz GW signal under the adopted dynamical and thermal assumptions.
The leading-order friction estimate motivates our assumption that bubble collisions dominate the GW signal.
To test how spatial temperature variations affect big bang nucleosynthesis (BBN), we start with a cosine perturbation of the initial temperatures and evolve it through nuclear burning.
We compare the resulting deuterium and helium abundances with observations for a benchmark point and variations of the dark-sector and reheating parameters.
We compare the predicted spectral shapes and amplitudes with PTA free-spectrum posteriors and examine how dark-sector and reheating parameters affect the relation between the relic abundance and gravitational waves.

Sections~\ref{sec:mod:model}--\ref{sec:asymdm} describe the model, initial conditions, phase transition, thermal evolution, GW spectrum and asymmetric relic abundance.
Section~\ref{sec:benchmark-results} presents the benchmark results and their dependence on the dark-sector and reheating parameters.
Section~\ref{sec:phenomenological-implications} presents the BBN comparisons.
We conclude in Section~\ref{sec:conclusions}.

\section{Dark sector model}
\label{sec:mod:model}
\subsection{Field content and tree-level potential}
The dark sector contains a real scalar $\phi$, a complex scalar $S=(H_D+iA_S)/\sqrt{2}$, and a Dirac fermion $\chi$ that serves as dark matter (DM).
All three fields are Standard Model singlets.
The dark sector Lagrangian is given by
\begin{equation}
\mathcal L_D=\frac12(\partial_\mu\phi)^2+|\partial_\mu S|^2+\overline\chi i\slashed\partial\chi-y_\chi\phi\overline\chi\chi-V_0(\phi,S),
\label{mod:lagrangian}
\end{equation}
where the tree-level scalar potential is given by~\cite{Wong:2023qon}
\begin{equation}
V_0(\phi,S)=\frac{\lambda_\phi}{4}\phi^4+\lambda_S|S|^4+\frac{\lambda_{\phi S}}2\phi^2|S|^2.
\label{mod:scalar-potential}
\end{equation}
The potential $V_0$ is classically conformal, containing only quartic interactions and no explicit mass scale.
The $\mathbb{Z}_2^{\rm chiral}$ symmetry in Table~\ref{mod:source-charges} allows the Yukawa interaction $y_\chi\phi\overline\chi\chi$ while forbidding a bare Dirac mass.

To generate the DM asymmetry, we also introduce an AD field $X$~\cite{Affleck:1984fy,Bell:2011tn,Cheung:2011if,Petraki:2013wwa}.
To make the chiral structure of the interaction explicit, we write $\chi=(\xi,\zeta^\dagger)^\top$, where $\xi$ and $\zeta$ are left-handed Weyl fields.
We include both charge-transfer interactions allowed by the symmetry assignments in Table~\ref{mod:source-charges},
\begin{equation}
\Delta\mathcal L_X=|\partial_\mu X|^2-V_X-\frac12\left(\lambda_\xi X^\dagger\xi\xi+\lambda_\zeta X\zeta\zeta+\mathrm{h.c.}\right),
\label{mod:source-interactions}
\end{equation}
where we adopt the quadratic limit of the AD field potential in Ref.~\cite{Lloyd-Stubbs:2020sed},
\begin{equation}
V_X=m_X^2|X|^2+\frac12\left(bX^2+\mathrm{h.c.}\right).
\label{mod:source-potential}
\end{equation}
The $bX^2$ term explicitly breaks the global $U(1)_D$ symmetry and splits the masses of the two real components of $X$.
We impose $m_X^2>|b|$ to ensure that both components have positive squared masses.

The symmetry assignments governing the allowed interactions and the stability of $\chi$ are summarized in Table~\ref{mod:source-charges}.
Since $X$ carries $U(1)_D$ charge two, the $bX^2$ term explicitly breaks the global $U(1)_D$ symmetry to its $\mathbb{Z}_4^D$ subgroup.
Under the generator of $\mathbb{Z}_4^D$, the bilinears $\xi\xi$ and $\zeta\zeta$ change sign, so vacuum Majorana masses are forbidden.
The order-two subgroup of $\mathbb{Z}_4^D$ sends $\chi\to-\chi$ while leaving all scalar and Standard Model fields invariant.
The remaining $\mathbb{Z}_2$ subgroup of $\mathbb{Z}_4^D$ ensures the stability of $\chi$, which is the lightest particle odd under this symmetry.

\begin{table}[!htbp]
\centering
\begin{tabular}{lccccc}
\toprule
Field & $\phi$ & $S$ & $\xi$ & $\zeta$ & $X$ \\
\midrule
$U(1)_D$ & $0$ & $0$ & $1$ & $-1$ & $2$ \\
$\mathbb Z_2^{\rm chiral}$ & $-1$ & $-1$ & $-1$ & $1$ & $1$ \\
$\mathbb Z_4^D$ & $1$ & $1$ & $i$ & $-i$ & $-1$ \\
\bottomrule
\end{tabular}
\caption{Dark-sector field content and symmetry assignments.
The $U(1)_D$ row gives charges.
The rows for the discrete symmetries give the phases acquired by each field under the corresponding generator.}
\label{mod:source-charges}
\end{table}

In addition to the interactions within the dark sector, we include the following dimension-six operators that reheat the Standard Model after the phase transition:
\begin{equation}
\mathcal L_{\rm reh}\supset-\frac{c_e y_e^{\rm SM}}{2\Lambda_e^2}\phi^2\overline L_eH_{\rm SM}e_R-\frac{c_{S\mu}y_\mu^{\rm SM}}{\sqrt2\Lambda_\mu^2}\phi S\overline L_\mu H_{\rm SM}\mu_R+\mathrm{h.c.},
\label{reh:operators}
\end{equation}
where $y_\ell^{\rm SM}v_{\rm EW}/\sqrt2=m_\ell$, and we take the Wilson coefficients $c_e, c_{S\mu} $ to be real.

\subsection{Radiative symmetry breaking and reheating channels}
\label{sec:portals}
To realize radiative symmetry breaking along the $\phi$ direction, we adopt the Gildener--Weinberg formalism~\cite{Coleman:1973jx,Gildener:1976ih}.
At the Gildener--Weinberg scale $\mu_{\rm GW}$, we impose $\lambda_\phi=0$ and $\lambda_S,\lambda_{\phi S}>0$, yielding a stable tree-level flat direction along the $\phi$ axis at $S=0$.
Radiative corrections lift this flat direction and can generate a symmetry-breaking minimum at $\phi=v$.
For general backgrounds, the scalar mass-squared eigenvalues follow from the Hessian of $V_0$ in the $(\phi,H_D,A_S)$ basis,
\begingroup
\setlength{\arraycolsep}{1pt}
\begin{equation}
\mathcal M^2=\begin{pmatrix}
3\lambda_\phi\phi^2+\frac12\lambda_{\phi S}(H_D^2+A_S^2)&\lambda_{\phi S}\phi H_D&\lambda_{\phi S}\phi A_S\\
\lambda_{\phi S}\phi H_D&\lambda_S(3H_D^2+A_S^2)+\frac12\lambda_{\phi S}\phi^2&2\lambda_SH_DA_S\\
\lambda_{\phi S}\phi A_S&2\lambda_SH_DA_S&\lambda_S(H_D^2+3A_S^2)+\frac12\lambda_{\phi S}\phi^2
\end{pmatrix}.
\label{mod:tree-matrix}
\end{equation}
\endgroup
On the $S=0$ axis, the squared tree-level scalar masses satisfy $m_{H_D,\mathrm{tree}}^2=m_{A_S,\mathrm{tree}}^2=\lambda_{\phi S}\phi^2/2$.
The fermion mass is $m_\chi(\phi)=y_\chi\phi$, giving $M_\chi=y_\chi v$ in the broken vacuum.
When no background is specified, $m_{H_D,\mathrm{tree}}$ and $m_{A_S,\mathrm{tree}}$ denote their values at $(v,0)$.
The one-loop Coleman--Weinberg potential is~\cite{Martin:2001vx,Jackiw:1974cv}
\begin{equation}
\begin{split}
V_{\rm CW}(\phi,H_D)={}\frac{1}{64\pi^2}\Bigg\{\sum_{b=1}^{3}m_b^4\left[\ln\frac{|m_b^2|}{\mu_{\rm GW}^2}-\frac32\right]-4m_\chi^4\left[\ln\frac{m_\chi^2}{\mu_{\rm GW}^2}-\frac32\right]\Bigg\},
\end{split}
\label{mod:one-loop-mass-sum}
\end{equation}
where $m_b^2$ are the eigenvalues of the tree-level scalar mass-squared matrix evaluated at $A_S=0$, and $m_\chi^2=y_\chi^2\phi^2$.
The vacuum-subtracted effective potential is
\begin{equation}
\Veff(\phi,H_D;0)=V_0(\phi,H_D)+V_{\rm CW}(\phi,H_D)-V_{\rm CW}(v,0).
\label{mod:two-field-vacuum}
\end{equation}
The condition $\left.\dd\Veff(\phi,0;0)/\dd\phi\right|_{\phi=v}=0$ gives
\begin{equation}
\ln\frac{\mu_{\rm GW}^2}{v^2}=\frac{\lambda_{\phi S}^2\ln(\lambda_{\phi S}/2)-8y_\chi^4\ln y_\chi^2}{\lambda_{\phi S}^2-8y_\chi^4}-1.
\label{mod:vacuum-matching-scale}
\end{equation}
For simplicity, we define
\begin{equation}
B\equiv\frac{m_{H_D,\mathrm{tree}}^4+m_{A_S,\mathrm{tree}}^4-4M_\chi^4}{64\pi^2v^4}=\frac{\lambda_{\phi S}^2-8y_\chi^4}{128\pi^2}.
\label{mod:cw-coefficient}
\end{equation}
The potential along the real-scalar direction can then be written as
\begin{equation}
\Veff(\phi,0;0)=B\phi^4\left[\ln\frac{\phi^2}{v^2}-\frac12\right]+\frac{Bv^4}{2}.
\label{mod:cw-compact}
\end{equation}
A radiative minimum requires $B>0$ and has $m_{\sigma,\mathrm{LO}}^2=8Bv^2$.
The vacuum-energy density difference is
\begin{equation}
\begin{split}
\Delta V_0\equiv\Veff(0,0;0)-\Veff(v,0;0)=\frac{Bv^4}{2}=\frac{m_{\sigma,\mathrm{LO}}^2v^2}{16}.
\end{split}
\label{mod:vacuum-energy-scale}
\end{equation}

At temperature $T_{\mathrm{true}}$ and fermion chemical potential $\mu_{\chi,\mathrm{true}}$, we approximate the thermal contributions to the broken-phase grand potential by those of massive ideal gases~\cite{Dolan:1973qd,Laine:2016hma}.
With $m_\sigma^2=m_{\sigma,\mathrm{LO}}^2$ fixed at its vacuum value and $\eta_{\chi,\mathrm{true}}=\mu_{\chi,\mathrm{true}}/T_{\mathrm{true}}$, the grand-potential density is
\begin{equation}
\begin{split}
V_{{\rm eff},\mathrm{true}}(\phi,T_{\mathrm{true}},\eta_{\chi,\mathrm{true}})={}&\Veff(\phi,0;0)+\frac{T_{\mathrm{true}}^4}{2\pi^2}\Biggl[2J_B\!\left(\frac{\lambda_{\phi S}\phi^2}{2T_{\mathrm{true}}^2}\right)+J_B\!\left(\frac{m_\sigma^2}{T_{\mathrm{true}}^2}\right)\\
&\hspace{4.2em}-4J_F\!\left(\frac{y_\chi^2\phi^2}{T_{\mathrm{true}}^2},\eta_{\chi,\mathrm{true}}\right)\Biggr],
\end{split}
\label{pt:phase-grand-potentials}
\end{equation}
where the bosonic and fermionic thermal functions $J_B$ and $J_F$ are defined in Equation~\eqref{mod:thermal-integrals}.
In Section~\ref{sec:cosmological-evolution}, we evaluate this potential at its broken-phase minimum to obtain the equation of state.

After symmetry breaking, we write $\phi=v+\sigma$.
The interactions in Eq.~\eqref{reh:operators} then give scalar couplings $-g_{\sigma e}\sigma\overline ee-g_{H_D\mu}H_D\overline\mu\mu$ and a pseudoscalar coupling $-ig_{A_S\mu}A_S\overline\mu\gamma_5\mu$, with
\begin{equation}
g_{\sigma e}=\frac{c_em_ev}{\Lambda_e^2},\qquad g_{H_D\mu}=g_{A_S\mu}=\frac{c_{S\mu}m_\mu v}{2\Lambda_\mu^2}.
\label{reh:couplings}
\end{equation}
Above the corresponding two-lepton thresholds, the visible decay widths are~\cite{Abdallah:2015ter}
\begin{equation}
\begin{split}
\Gamma_{\sigma\to e^+e^-}&=\frac{g_{\sigma e}^2m_\sigma}{8\pi}\left(1-\frac{4m_e^2}{m_\sigma^2}\right)^{3/2},\\
\Gamma_{H_D\to\mu^+\mu^-}&=\frac{g_{H_D\mu}^2m_{H_D}}{8\pi}\left(1-\frac{4m_\mu^2}{m_{H_D}^2}\right)^{3/2},\\
\Gamma_{A_S\to\mu^+\mu^-}&=\frac{g_{A_S\mu}^2m_{A_S}}{8\pi}\left(1-\frac{4m_\mu^2}{m_{A_S}^2}\right)^{1/2}.
\end{split}
\label{reh:visible-widths}
\end{equation}

\subsection{Affleck--Dine dynamics and initial conditions}
\label{sec:asymmetry-origin}
Writing $X=Re^{-i\theta}/\sqrt2$, we choose the phase so that $b$ is real and positive.
We specify an initially uncharged AD condensate during the radiation-dominated era, when $H\gg m_X$, and impose equal initial temperatures for the dark and visible sectors,
\begin{equation}
T_{\mathrm{init}}\equiv T_\gamma(t_{\mathrm{init}})=T_D(t_{\mathrm{init}}),\qquad
\dot\theta_{\mathrm{init}}=0.
\label{adm:initial-temperature-equality}
\end{equation}
Here $\theta_{\mathrm{init}}=\theta(t_{\mathrm{init}})$ is the initial phase.
The temperature equality is an imposed boundary condition.
We assume negligible energy exchange between the dark thermal bath at temperature $T_D$ and the Standard Model thermal bath at temperature $T_\gamma$ during the AD evolution, rather than deriving it by extrapolating the MeV-scale reheating interactions to high temperature.
The AD condensate transfers energy and charge to the dark bath through its decays.
The initial radial field and its small radial velocity are fixed by matching the regular radiation-era solution to the oscillation-envelope normalization defined below.
We denote the frozen radial amplitude before oscillations begin by $R_{\rm frozen}$.
The same cosmic-time origin and scale factor, normalized by $a_p=1$, are used throughout.

The condensate starts oscillating at $H\sim m_X$.
With $X=(x+iy)/\sqrt2$, the squared masses of the two real components are $m_x^2=m_X^2+b$ and $m_y^2=m_X^2-b$, with $\Delta m=m_x-m_y\simeq b/m_X$.
Neglecting medium corrections and the small mass splitting in the decay kinematics, we obtain the total tree-level width into massless fermions
\begin{equation}
\Gamma_X=\Gamma_\xi+\Gamma_\zeta=\frac{m_X}{32\pi}\lambda_{\rm eff}^2,\qquad
\lambda_{\rm eff}^2\equiv|\lambda_\xi|^2+|\lambda_\zeta|^2,
\label{adm:total-width}
\end{equation}
where $\lambda_{\rm eff}\geq0$ and $\Gamma_\xi$ and $\Gamma_\zeta$ are the two partial widths.

We define the $U(1)_D$ charge density of the AD condensate as $n_X=2R^2\dot\theta$ and the net fermion number density as $d=n_\chi-n_{\overline\chi}$.
The perturbative AD evolution obeys
\begin{equation}
\begin{split}
&\ddot X+(3H+\Gamma_X)\dot X+m_X^2X+bX^*={}0,\\
&\dot n_X+(3H+\Gamma_X)n_X={}-4\operatorname{Im}(bX^2),\\
&\dot d+3Hd={}\Gamma_Xn_X.
\end{split}
\label{adm:source-evolution}
\end{equation}
For real positive $b$, the source term is $-4\operatorname{Im}(bX^2)=2bR^2\sin(2\theta)$, so an initially uncharged condensate develops a charge asymmetry if $\sin(2\theta_{\mathrm{init}})\neq0$~\cite{Affleck:1984fy,Dine:1995kz,Lloyd-Stubbs:2020sed}.
The decay transfers charge, while only the $b$ term changes $a^3(n_X+d)$.
After we average over its oscillations, the condensate behaves as pressureless matter and heats a relativistic dark sector according to~\cite{Adshead:2019uwj}
\begin{equation}
\begin{split}
&\dot\rho_X+3H\rho_X={}-\Gamma_X\rho_X,\qquad\\
&\dot\rho_D+4H\rho_D=\Gamma_X\rho_X,\\
&H^2={}\frac{\rho_X+\rho_D+\rho_{\rm SM}+\Delta V_0}{3\Mpl^2},
\end{split}
\label{adm:source-energy}
\end{equation}
where $H=\dot a/a$ is the Hubble rate and $\Mpl$ is the reduced Planck mass.
Here $\rho_D$ denotes the energy density of relativistic dark-sector particles, excluding the condensate energy $\rho_X$ and the false-vacuum energy $\Delta V_0$.
We assume rapid kinetic and chemical equilibration among the relativistic dark-sector particles throughout the AD decay, with $\mu_\xi=\mu_\chi$ and $\mu_\zeta=-\mu_\chi$.
The dark-sector temperature $T_D$ and chemical-potential ratio $\eta_\chi=\mu_\chi/T_D$ then follow from the massless Bose--Einstein and Fermi--Dirac equilibrium distributions, with vanishing scalar chemical potentials~\cite{Laine:2016hma},
\begin{equation}
d=\frac{T_D^3}{3}\left(\eta_\chi+\frac{\eta_\chi^3}{\pi^2}\right),
\qquad
\rho_D=T_D^4\left(\frac{g_{*,D}\pi^2}{30}+\frac{\eta_\chi^2}{2}
+\frac{\eta_\chi^4}{4\pi^2}\right),
\label{adm:source-bath}
\end{equation}
where $g_{*,D}=1+2+(7/8)\times4=6.5$ is the effective number of relativistic degrees of freedom of dark sector particles.
These dark-bath expressions neglect thermal masses and interaction corrections.
Within the approximation of massless decay products in a common dark-sector thermal bath, charge transfer and heating depend on the two Yukawa couplings only through $\lambda_{\rm eff}$.
At fixed $\lambda_{\rm eff}$ and other inputs, any partition between $|\lambda_\xi|^2$ and $|\lambda_\zeta|^2$ gives the same source evolution and hence the same relic and GW predictions within the stated approximation.

After oscillations begin, we define the slowly varying oscillation amplitude of the AD field $X$ in the radial normalization $R=\sqrt2|X|$ by $\mathcal A_X^2\equiv2\langle R^2\rangle_{\rm osc}$, where the average removes the rapid oscillations without averaging over the slow beat evolution.
In the oscillation-averaged approximation, this amplitude evolves as $\mathcal A_X\simeq R_{\rm env}e^{-\Gamma_Xt/2}$, where $R_{\rm env}=R_{\rm env,ref}(a_{\rm ref}/a)^{3/2}$.
We separate the decay factor $e^{-\Gamma_Xt/2}$ so that $R_{\rm env}$ describes the decrease in amplitude due to expansion.
The physical amplitude $\mathcal A_X$ includes both effects.
Here $a_{\rm ref}$ is the scale factor at the reference epoch at which $T_\gamma=T_{\gamma,\rm ref}$.
The envelope normalization $R_{\rm env,ref}$ differs from the initial field amplitude.
In the regime $m_X\gg H,\Gamma_X,b/m_X$, define ${\cal N}_X=\tfrac12m_XR_{\rm env,ref}^2a_{\rm ref}^3$.
The leading charge-transfer solution follows from the oscillation-averaged quadratic-potential result of Ref.~\cite{Lloyd-Stubbs:2020sed}, with the $U(1)_D$ charge of $X$ equal to two,
\begin{equation}
\begin{split}
Q_X(t)&\equiv a^3n_X\simeq
2{\cal N}_X\sin(2\theta_{\mathrm{init}})e^{-\Gamma_Xt}\sin(\Delta m\,t),\\
Q_\chi(\infty)&=
\int_0^\infty\dd t\,\Gamma_XQ_X(t)
\simeq2{\cal N}_X\sin(2\theta_{\mathrm{init}})\frac{r_X}{1+r_X^2},
\qquad
r_X=\frac{b}{m_X\Gamma_X}.
\end{split}
\label{adm:early-source-inputs}
\end{equation}
The leading analytic expression neglects decay and charge transfer during the onset of oscillations.
The leading efficiency $\epsilon_X=r_X/(1+r_X^2)$ is largest at $r_X=1$, but the final abundance also depends on expansion and entropy production.

In the numerical calculation, we follow the regular radiation-era solution through the onset of oscillations at $H\sim m_X$ before averaging over the oscillations.
We carry forward the charge $Q_{\chi,m}$ already transferred and the beat phase $\varphi_m$ at the matching time $t_m$, and define $u_m=\Gamma_Xt_m$.
Integrating the charge-transfer equation in Eq.~\eqref{adm:source-evolution} using the oscillation-averaged condensate charge gives
\begin{equation}
Q_\chi(\infty)\simeq Q_{\chi,m}+2{\cal N}_X\sin(2\theta_{\mathrm{init}})e^{-u_m}\frac{\sin\varphi_m+r_X\cos\varphi_m}{1+r_X^2}.
\label{adm:source-matching-charge}
\end{equation}
The later beat phase is $\varphi_m+r_X(u-u_m)$, with $u=\Gamma_Xt$.
After the AD condensate decays, $Q_\chi=a^3d$ and ${\cal E}_D=a^4\rho_D$ are conserved in the free-radiation approximation.
Consequently, for small $|\eta_\chi|$, the chemical-potential ratio remains constant and is given by
\begin{equation}
\eta_\chi\simeq3\left(\frac{13\pi^2}{60}\right)^{3/4}\frac{d}{\rho_D^{3/4}}.
\label{adm:late-chemical-potential}
\end{equation}

For the numerical calculations, we adopt $m_X=10\GeV$ and $\Gamma_X=7.412\times10^{-19}\GeV$, corresponding to $\lambda_{\rm eff}=2.730\times10^{-9}$ through Eq.~\eqref{adm:total-width}.
We choose
\begin{equation}
T_{\gamma,\rm ref}=3\times10^9\GeV,\qquad
R_{\rm env,ref}=10^{14}\GeV,\qquad
\theta_{\mathrm{init}}=\frac\pi8.
\label{adm:early-envelope}
\end{equation}
We start the numerical evolution at $H(t_{\mathrm{init}})/m_X=1000$.

To illustrate the evolution in the following sections, we introduce benchmark B1 with the parameters in Table~\ref{tab:benchmark-summary} and the AD parameters specified above.
Only $r_X$ is adjusted to give $\Omega_\chi h^2=0.12$ within a relative tolerance of $0.002$, using instantaneous neutrino decoupling.
For the reheating interactions of Eq.~\eqref{reh:operators}, we take
\begin{equation}
c_e=c_{S\mu}=0.1,\qquad \Lambda_e=200.4\GeV,\qquad \Lambda_\mu=2953\GeV.
\label{reh:portal-inputs}
\end{equation}

The AD initial conditions above give $T_{\rm init}=8.312\times10^{10}\GeV$, while matching the regular radiation-era solution to the envelope normalization gives $R_{\rm frozen}=5.670\times10^{13}\GeV$ for B1.
At $T_\gamma=100\GeV$, the initial temperature equality would give $T_D^{(0)}=98.09\GeV$ without AD heating.
For B1 the oscillation amplitude with the decay factor removed at that epoch is $R_{\rm env}=591.3\GeV$.

\FloatBarrier

\section{Phase transition dynamics}
\label{sec:transition}
Each thermal bounce is evaluated at fixed symmetric-phase temperature and chemical potential, which evolve along the cosmological history.
For the benchmark histories, we use the single-channel rate for thermal nucleation from the false vacuum at the origin toward $\phi=+v$, evaluated in the unbiased potential.\footnote{A small explicit breaking of $\mathbb Z_2^{\rm chiral}$ could favor nucleation toward $+v$ over $-v$, as illustrated for analogous transitions in singlet scalar models~\cite{Cline:2021iff,Liu:2023sey}.
The formation and evolution of domain walls are beyond the scope of this work.}

\subsection{Thermal nucleation rate in a 3D effective theory}
\label{sec:thermal-potentials}
Daisy resummation incorporates leading thermal screening effects, but using the resummed effective potential alone does not systematically organize the fluctuation and gradient corrections to the nucleation rate.
When the nucleation scale is well below $\pi T_D$, dimensional reduction provides a three-dimensional effective field theory (EFT) for the long-wavelength bosonic zero modes.
We use this framework to organize hard-mode matching and soft corrections and to avoid double counting between the bounce action and fluctuation determinants~\cite{Gould:2021ccf}.

We calculate the thermal nucleation rate using the NLOdet prescription~\cite{Kierkla:2026bnm}.
First, integrating out fermions and nonzero bosonic Matsubara modes gives a three-dimensional thermal EFT~\cite{Kajantie:1995dw}.
To determine the EFT parameters, we use DRalgo~\cite{Ekstedt:2022bff} for the one-loop renormalization-group evolution of the four-dimensional couplings and for matching the scalar masses and quartic couplings at two-loop and one-loop order, respectively, at zero chemical potential.
We impose boundary conditions at $\mu_{\rm GW}$ and choose the renormalization scales $\bar\mu_4=\pi T_D$ and $\bar\mu_3=T_D$.
At leading order, the thermal mass corrections at zero chemical potential are~\cite{Weinberg:1974hy}
\begin{equation}
\Pi_\phi^{(0)}=\TD^2\left(\frac{\lambda_\phi}{4}+\frac{\lambda_{\phi S}}{12}+\frac{y_\chi^2}{6}\right),\qquad \Pi_H=\Pi_A=\TD^2\left(\frac{\lambda_S}{3}+\frac{\lambda_{\phi S}}{24}\right).
\label{mod:thermal-masses}
\end{equation}
In the symmetric phase ($\phi=0$), the massless fermion gives a chemical-potential-dependent contribution to the scalar thermal mass.
Including this contribution, we obtain
\begin{equation}
\Pi_\phi=\Pi_\phi^{(0)}+\frac{y_\chi^2\mu_\chi^2}{2\pi^2}=\TD^2\left[\frac{\lambda_\phi}{4}+\frac{\lambda_{\phi S}}{12}+y_\chi^2\left(\frac16+\frac{\eta_\chi^2}{2\pi^2}\right)\right].
\label{mod:finite-density-mass}
\end{equation}
Appendix~\ref{app:density-matching} gives the additional density corrections to field normalization and quartic couplings at one loop and to scalar masses at two loops.

Using these matched parameters, we construct the leading-order potential including the local one-loop contribution of the soft $S$ modes.
To simplify the bounce equations, we introduce the dimensionless variables $x=\phi_3/\sqrt{T_D}$, $\rho=r_bT_D$, $a_i=m_{i,3}^2/T_D^2$ and $\ell_i=\lambda_{i,3}/T_D$, where $\phi_3$ is the canonically normalized field and $r_b$ is the radial coordinate.
Expressed as functions of $x$, the leading-order three-dimensional potential and the screened $S$ mass $M_S(x)$ in units of $T_D$ are
\begin{equation}
\begin{split}
V_{\rm 3d}^{\rm LO}(x)&=T_D^3\left\{\frac12a_\phi x^2+\frac14\ell_\phi x^4-\frac{1}{6\pi}\left[M_S^3(x)-M_S^3(0)\right]\right\},\\
M_S^2(x)&=a_S+\frac12\ell_{\phi S}x^2.
\end{split}
\label{pt:nlodet-potential}
\end{equation}
This potential determines the spherical LO bounce $x_0$ and its action $S_0$ through~\cite{Kierkla:2026bnm}
\begin{equation}
\begin{split}
S_0&=4\pi\int_0^\infty\dd\rho\,\rho^2\left[\frac12x_0'^2+\frac{V_{\rm 3d}^{\rm LO}(x_0)}{T_D^3}\right],\\
x_0''+\frac{2}{\rho}x_0'&=\frac{\partial_x V_{\rm 3d}^{\rm LO}(x_0)}{T_D^3},\qquad x_0'(0)=0,\quad x_0(\infty)=0.
\end{split}
\label{pt:nlodet-action}
\end{equation}
Here primes on $x_0$ denote derivatives with respect to $\rho$, while $\partial_x$ acts at fixed $T_D$ and $\eta_\chi$.

We evaluate the soft two-loop potential correction on this fixed LO bounce $x_0$ to obtain the corresponding correction to the action~\cite{Gould:2021ccf,Hirvonen:2021zej}.
For this potential correction, we assume that the masses of the soft $S$ modes are much larger than that of the soft $\phi$ mode.
We therefore neglect the $\phi$ mass in the mixed scalar diagram and omit loops from $\phi$ self-interactions.
The resulting potential and action corrections are
\begin{equation}
\begin{split}
\Delta V_{\rm 3d,heavy}^{(2)}(x)&=T_D^3\left\{\frac{\ell_SM_S^2(x)}{8\pi^2}-\frac{\ell_{\phi S}^2x^2}{32\pi^2}\left[\ln\frac{\bar\mu_3/T_D}{2M_S(x)}+\frac12\right]-(x\to0)\right\},\\
\Delta S_{2,\rm heavy}&=4\pi\int\dd\rho\,\rho^2\frac{\Delta V_{\rm 3d,heavy}^{(2)}(x_0)}{T_D^3}.
\end{split}
\label{pt:heavy-soft-correction}
\end{equation}
We also calculate the one-loop corrections from small fluctuations of $\phi$ and $S$ around the LO bounce $x_0(\rho)$.
We use BubbleDet~\cite{Ekstedt:2023sqc} to evaluate the functional determinants with the full radial dependence of the mass terms, without a derivative expansion.
The local cubic $S$-loop term is already included in $S_0$ through $V_{\rm 3d}^{\rm LO}$ and also appears in the logarithm of the full $S$ determinant.
We subtract this common term and denote the remaining one-loop $S$ contribution to the effective action by $\Delta E_S$.
As for the two-loop correction, we evaluate the determinant on the fixed LO bounce $x_0(\rho)$.
With $\widehat O_{S,b}=-\nabla_\rho^2+M_S^2(x_0)$ and $\widehat O_{S,\mathrm{false}}=-\nabla_\rho^2+M_S^2(0)$, this gives
\begin{equation}
\begin{split}
\Delta E_S&=\ln\frac{\det\widehat O_{S,b}}{\det\widehat O_{S,\mathrm{false}}}-4\pi\int\dd\rho\,\rho^2\frac{V_{{\rm 3d},S,\rm loc}(x_0)}{T_D^3},\\
V_{{\rm 3d},S,\rm loc}(x)&=-\frac{T_D^3}{6\pi}\left[M_S^3(x)-M_S^3(0)\right].
\end{split}
\label{pt:spectator-local-subtraction}
\end{equation}
The full determinant also includes the spatial-gradient effects of this one-loop $S$ contribution, so no separate $S$-loop kinetic correction is added.

The LO potential contains no local one-loop contribution from the soft $\phi$ mode, so the $\phi$ determinant requires no analogous subtraction.
Its fluctuation operators are $\widehat O_{\phi,b}=-\nabla_\rho^2+T_D^{-3}\partial_x^2V_{\rm 3d}^{\rm LO}(x_0)$ and $\widehat O_{\phi,\mathrm{false}}=-\nabla_\rho^2+T_D^{-3}\partial_x^2V_{\rm 3d}^{\rm LO}(0)$, with negative bounce eigenvalue $\widehat\lambda_-$.
Including the undamped growth factor $A_{\rm dyn}/T_D=\sqrt{|\widehat\lambda_-|}/(2\pi)$~\cite{Ekstedt:2021kyx}, the real-scalar contribution is
\begin{equation}
S_{1,\phi}^{\rm thermal}=\frac12\ln\left|\frac{\det'\widehat O_{\phi,b}}{\det\widehat O_{\phi,\mathrm{false}}}\right|-\frac32\ln\frac{S_0}{2\pi}-\ln\frac{\sqrt{|\widehat\lambda_-|}}{2\pi}.
\label{pt:radial-normalization}
\end{equation}
Here the prime removes the three translation zero modes, whose normalization gives the second term, while the last term accounts for the undamped growth factor.
The determinant includes all partial waves, with its negative mode included inside the absolute value.

Combining these one-loop contributions with the LO action and the soft two-loop correction gives the thermal nucleation rate
\begin{equation}
\Gamma_{\rm NLOdet}=T_D^4\exp\left[-S_0-S_{1,\phi}^{\rm thermal}-\Delta E_S-\Delta S_{2,\rm heavy}\right].
\label{pt:nlodet-rate}
\end{equation}
The high-temperature expansion and heavy-$S$ approximation are model-dependent assumptions, and plasma dissipation is omitted, so this prescription does not give the complete physical rate at relative NLO.
Section~\ref{sec:rate-diagnostics} examines these assumptions for B1.

The bulk equations of state used in the cosmological evolution are specified in Section~\ref{sec:cosmological-evolution}.

\subsection{Nucleation, completion and bubble separation}
To follow the phase transition, we integrate the thermal nucleation rate $\Gamma_{\rm nuc}$ from $t_{\mathrm{nuc,init}}$, with bubbles expanding at wall speed $v_w$.
We set $v_w=1$, $\Gamma_{\rm nuc}=\Gamma_{\rm NLOdet}$ and $a_p=1$.
For homogeneous thermal nucleation, the probability that a point remains in the false vacuum is~\cite{Guth:1982pn,Turner:1992tz}
\begin{equation}
\begin{split}
P_{\mathrm{false}}(t)={}&e^{-I(t)},\\
I(t)={}&\frac{4\pi}{3}\int_{t_{\mathrm{nuc,init}}}^{t}\dd t'\,\Gamma_{\mathrm{nuc}}(t')a^3(t')\left[\int_{t'}^{t}\frac{v_w(t'')\dd t''}{a(t'')}\right]^3.
\end{split}
\label{pt:percolation}
\end{equation}
The fraction of the volume in the true vacuum is $P_{\rm true}=1-P_{\rm false}$.
Percolation is defined by a true-vacuum fraction $P_{\rm true}(t_p)\simeq0.2896$~\cite{Matuszak:2026xsz}, corresponding to $I(t_p)=-\ln[1-P_{\rm true}(t_p)]\simeq0.34$.
The nucleation time $t_n$ instead follows from the cumulative nucleation count $\mathcal N_{\rm nuc}$ through~\cite{Athron:2022mmm}
\begin{equation}
\mathcal N_{\rm nuc}(t)=\frac{4\pi v_w^3}{3}\int_{t_{\mathrm{nuc,init}}}^{t}\dd t'\,\frac{\Gamma_{\rm nuc}(t')P_{\mathrm{false}}(t')}{H(t')^3},\qquad \mathcal N_{\rm nuc}(t_n)=1.
\label{pt:nucleation-criterion}
\end{equation}
During vacuum domination, the false-vacuum energy sets the Hubble scale $H_{\rm vac}^2=\Delta V_0/(3\Mpl^2)$.
For the characteristic time and length scales, we use the Hubble rate at percolation, $H_*=H(t_p)$, calculated from the total energy density of the two phases, bubble walls and visible radiation as described in Section~\ref{sec:reh:reheating}.

As the symmetric-phase bath evolves, we evaluate the rate from its values at zero chemical potential and the tabulated reference ratio $\eta_{\rm ref}$.
The physical initial chemical-potential ratio need not equal $\eta_{\rm ref}$.
Within the validated temperature range, the rate is evaluated using
\begin{equation}
\ln\frac{\Gamma(T_{\mathrm{false}},\eta_{\chi,\mathrm{false}})}{\Gamma(T_{\mathrm{false}},\eta_{\rm ref})}=\left[1-\left(\frac{\eta_{\chi,\mathrm{false}}}{\eta_{\rm ref}}\right)^2\right] \ln\frac{\Gamma(T_{\mathrm{false}},0)}{\Gamma(T_{\mathrm{false}},\eta_{\rm ref})}.
\label{pt:rate-density-interpolation}
\end{equation}
The logarithmic rate ratio is interpolated in temperature.
Calculations at intermediate chemical potentials test this quadratic approximation to $\ln\Gamma$ at fixed temperature.

Completion requires the physical false-vacuum volume to decrease~\cite{Ellis:2018mja}, so
\begin{equation}
\frac{\dd\ln(a^3P_{\mathrm{false}})}{\dd t}=3H-\dot I<0.
\label{pt:completion}
\end{equation}
We define the numerical completion time $t_{\mathrm{comp}}$ by $P_{\rm false}(t_{\mathrm{comp}})=10^{-8}$.
The subscript $\mathrm{comp}$ refers to this time; $T_{\mathrm{false},\mathrm{comp}}$ is the symmetric-phase temperature immediately before the residual symmetric phase is removed.
With $N=\ln(a/a_p)$, the completed histories also satisfy $\dd I/\dd N>3$ from percolation to completion.

To obtain the characteristic length scale for bubble collisions, we calculate the bubble number density $n_{\rm bubble}$ at percolation and the corresponding mean bubble separation $R_*$~\cite{Matuszak:2026xsz},
\begin{equation}
n_{\rm bubble}(t_p)=\frac{1}{a_p^3}\int_{t_{\mathrm{nuc,init}}}^{t_p}\dd t\,a^3(t)P_{\mathrm{false}}(t)\Gamma_{\rm nuc}(t),\qquad R_*=n_{\rm bubble}(t_p)^{-1/3}.
\label{pt:bubble-density}
\end{equation}
We define the corresponding inverse-duration parameter $\beta_R$ using a reference true-vacuum fraction $f_{\rm ref}$ and normalization $C_R$~\cite{Matuszak:2026xsz},
\begin{equation}
H_*R_* = \frac{C_R}{\beta_R/H_*},\qquad C_R=\left(\frac{8\pi}{f_{\rm ref}}\right)^{1/3},\qquad f_{\rm ref}\simeq0.2896.
\label{pt:beta_separation}
\end{equation}
The value of $f_{\rm ref}$ is set by this percolation threshold, while the numerical evolution adopts the rounded condition $I(t_p)=0.34$.
For expanding or non-exponential histories, $\beta_R$ need not equal a local action or rate slope.
For comparison with local slopes, define the cooling rate $q_T=-\dd\ln T_D/\dd\ln a$ and use $S_b=S_0$.
At fixed chemical-potential ratio, we evaluate the action slope $\beta_{S_b}/H=q_TT_D\dd S_b/\dd T_D$ and rate slope $\beta_\Gamma/H=-q_TT_D\dd\ln\Gamma/\dd T_D$.
A changing $\eta_\chi$ also contributes to the total rate slope.

\begin{figure}[!htbp]
\centering
\includegraphics[width=0.75\textwidth]{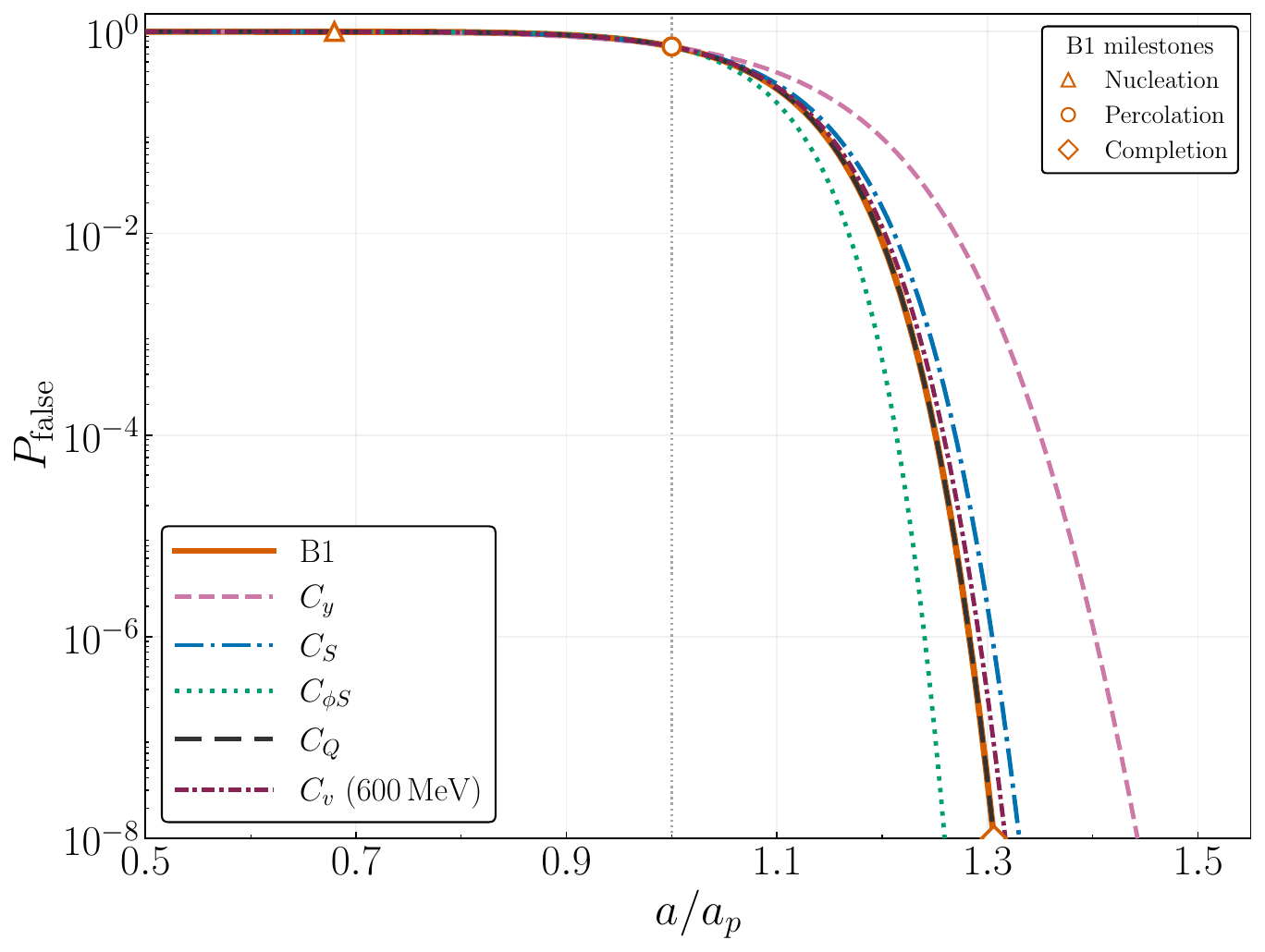}
\caption{False-vacuum fractions calculated with the NLOdet rate.
The curves compare B1 with the five dark-sector benchmarks in Table~\ref{tab:benchmark-summary}.
Each curve uses its own percolation scale factor $a_p$.
The triangle, circle and diamond mark nucleation, percolation and numerical completion.
The curves are shown through numerical completion at $P_{\mathrm{false}}=10^{-8}$.}
\label{fig:b1-transition}
\label{fig:benchmark-transition-milestones}
\end{figure}

For B1, $H_*R_*=0.3075$, and nucleation occurs at $a_n/a_p=0.6794$ and $T_{\rm false,n}=0.8094\MeV$.
Percolation occurs at $T_{\rm false,p}=0.5499\MeV$.
Numerical completion occurs at $a_{\mathrm{comp}}/a_p=1.306$ and $T_{\mathrm{false},\mathrm{comp}}=0.4848\MeV$.
Heat transfer makes the temperature evolution non-monotonic, so a temperature ratio cannot replace the integrated expansion factor.
Figure~\ref{fig:benchmark-transition-milestones} shows these stages and compares the benchmarks with varied dark-sector parameters.

\FloatBarrier

\section{Cosmological evolution}
\label{sec:cosmological-evolution}
To evolve the dark-sector temperatures and chemical potentials, we derive the required thermodynamic quantities from the effective potential of each phase.
Let $V_{{\rm eff},i}$ denote the effective potential evaluated at the local minimum corresponding to phase $i$, at temperature $T_i$ and chemical-potential ratio $\eta_\chi=\mu_i/T_i$.
The pressure $p_i$, fermion asymmetry $d_i$ and energy density $\rho_i$ follow from the thermodynamic relations~\cite{Laine:2016hma}
\begin{equation}
p_i=-V_{{\rm eff},i},\qquad d_i=-\frac{1}{T_i}\frac{\partial V_{{\rm eff},i}}{\partial\eta_\chi},\qquad \rho_i=V_{{\rm eff},i}-T_i\left.\frac{\partial V_{{\rm eff},i}}{\partial T_i}\right|_{\eta_\chi}.
\label{pt:phase-thermodynamics}
\end{equation}
The temperature derivative at fixed $\eta_\chi$ includes chemical energy.
The effective potentials in the false and broken phases share the zero-temperature vacuum normalization $\Veff(v,0;0)=0$ and $\Veff(0,0;0)=\Delta V_0$ of Eq.~\eqref{mod:cw-compact}.
The symmetric-phase potential in Eq.~\eqref{pt:false-grand-potential} includes interaction corrections, while the broken-phase potential in Eq.~\eqref{pt:phase-grand-potentials} combines the radiative vacuum potential with massive ideal-gas contributions.

\subsection{Initial conditions for the phase transition}
\label{sec:source-phase-matching}
The AD evolution determines the dark-sector radiation energy density $\rho_D$ and fermion asymmetry $d$ after condensate decay.
We use these densities to initialize the phase-transition calculation at a later matching epoch, whose scale factor we denote by $a_{\mathrm{ph,init}}$.
This epoch is distinct from the AD initial time $t_{\mathrm{init}}$ defined in Eq.~\eqref{adm:initial-temperature-equality}.
The matching scale factor is fixed in the convention used for the AD evolution; we subsequently rescale the full history to $a_p=1$.

Writing Eq.~\eqref{adm:source-bath} as $\rho_D=c_D(\eta_\chi)T_D^4$, we define the dimensionless energy coefficient
\begin{equation}
c_D(\eta)\equiv\frac{g_{*,D}\pi^2}{30}+\frac{\eta^2}{2}+\frac{\eta^4}{4\pi^2}.
\label{adm:source-energy-coefficient}
\end{equation}
For the initial conditions, we use $c_D(0)$ in place of $c_D(\eta_{\rm ref})$ at the same dark temperature, where $\eta_{\rm ref}$ is the fixed chemical-potential ratio used to tabulate the nucleation rate.
The initial radiation energy density is therefore rescaled by the factor $\mathcal M_D\equiv c_D(0)/c_D(\eta_{\rm ref})$.
We hold $\mathcal M_D$ fixed for each microscopic parameter set as the AD source is varied.
At this matching epoch, the initial symmetric-phase energy and charge densities are
\begin{equation}
\rho_{\mathrm{false},\mathrm{init}}=\Delta V_0+\mathcal M_D\rho_D(a_{\mathrm{ph,init}}),\qquad d_{\mathrm{ph,init}}=d(a_{\mathrm{ph,init}}).
\label{adm:phase-source-normalization}
\end{equation}
The thermodynamic relations in Eq.~\eqref{pt:phase-thermodynamics} then determine the initial symmetric-phase temperature and chemical potential from these energy and charge densities.
For B1, this matching epoch corresponds to $T_\gamma=12.06\MeV$.
Relative to the radiation energy density $\rho_D(a_{\mathrm{ph,init}})$ obtained from the AD evolution, the initial symmetric-phase radiation energy density is reduced by the fraction $1-\mathcal M_D=2.824\times10^{-4}$, while the charge density is unchanged.

\subsection{Energy exchange and visible-sector reheating}
\label{sec:reh:reheating}
We assume local thermal equilibrium within each phase and a spatially uniform fermion asymmetry $d=Q_\chi a^{-3}$.
The rates $\mathcal C_{E,\mathrm{false}}$ and $\mathcal C_{E,\mathrm{true}}$ denote net energy transfer to the visible sector per unit local phase volume and unit time.
We use $g_e=g_{\sigma e}$, $g_\mu=g_{H_D\mu}=g_{A_S\mu}$ and $\kappa_\mu=g_\mu/v$.
After electroweak symmetry breaking, the electron operator in Eq.~\eqref{reh:operators} gives the interaction $-g_e\phi^2\bar e e/(2v)$ in the symmetric phase of the dark sector.
We compute the energy transfer from $\phi\phi\leftrightarrow e^+e^-$ and $\phi e^\pm\leftrightarrow\phi e^\pm$ at fixed $g_e$, neglecting external masses in the kinematics.
The collision integrals use Maxwell--Boltzmann distributions per spin state at $T_{\rm false}$ and $T_\gamma$, zero electron chemical potential, and no final-state statistical factors~\cite{Adshead:2019uwj,EscuderoAbenza:2020cmq}.
Weighting annihilation by the pair energy, with an identical-pair factor $1/2$, and both $e^\pm$ scattering channels by the electron energy gain yields
\begin{equation}
\mathcal C_{E,\mathrm{false},e}=\frac{g_e^2}{8\pi^5v^2}\left[3(T_{\mathrm{false}}^7-T_\gamma^7)+2T_{\mathrm{false}}^3T_\gamma^3(T_{\mathrm{false}}-T_\gamma)\right].
\label{reh:false-transfer}
\end{equation}
The two terms describe annihilation and elastic scattering, respectively~\cite{Du:2021idh}.
The benchmark histories include this electron rate; the reheating-coupling comparisons also include
\begin{equation}
\mathcal L_{\mathrm{false},\mu}=-\kappa_\mu\phi H_D\overline\mu\mu-i\kappa_\mu\phi A_S\overline\mu\gamma_5\mu.
\label{reh:false-muon-operator}
\end{equation}
The leading-order integrals for $\phi H_D,\phi A_S\leftrightarrow\mu^+\mu^-$ and crossed scattering use the finite muon mass $m_\mu$, treat dark scalars as massless, and use the same statistical approximations at separate bath temperatures~\cite{Adshead:2019uwj}.
Their massless-muon limit is Eq.~\eqref{reh:false-transfer} with $g_e^2\to4g_\mu^2$.
At low visible temperature, scattering has single-muon Boltzmann suppression, whereas pair production has a two-muon threshold.
We sum the included lepton contributions to obtain $\mathcal C_{E,\mathrm{false}}$.

In the broken phase and the later homogeneous bath, scalar decays and inverse decays give the net energy transfer rate $\mathcal C_E$, positive for visible-sector heating.
For masses $m_i$, visible widths $\Gamma_{i\to\mathrm{SM}}$ and modified Bessel functions $K_1,K_2$, the energy transfer rate, equilibrium number density and decay reaction density are~\cite{EscuderoAbenza:2020cmq}
\begin{equation}
\begin{split}
\mathcal C_E={}&\sum_{i=\sigma,H_D,A_S}m_i\Gamma_{i\to\mathrm{SM}}\left[n_i^{\rm MB}(T_D)-n_i^{\rm MB}(T_\gamma)\right],\\
n_i^{\rm MB}(T)={}&\frac{m_i^2T}{2\pi^2}K_2(m_i/T),\\
\langle\Gamma_{{\rm dec},i}^{\rm MB}(T)\rangle={}&n_i^{\rm MB}(T)\Gamma_{i\to\mathrm{SM}}\frac{K_1(m_i/T)}{K_2(m_i/T)}.
\end{split}
\label{reh:transfer-moment}
\end{equation}
Setting $T_D=T_{\rm true}$ gives $\mathcal C_{E,\mathrm{true}}$.
The particle energy cancels time dilation in $\mathcal C_E$, whereas $\langle\Gamma_{{\rm dec},i}^{\rm MB}\rangle$ counts decays per unit volume and time.
We calculate the energy transfer rate using Maxwell--Boltzmann distributions for the decaying particles and neglect final-state statistical factors.
The scalar equation of state uses Bose--Einstein distributions.
The resulting uncertainty in the final entropy has not been quantified.

Phase conversion also deposits energy in ultrarelativistic bubble walls, with density $\rho_w$ and pressure $\rho_w/3$~\cite{Ellis:2019oqb}.
In the accelerating-wall limit of the thin-wall energy budget, the fraction of $\Delta V_0$ entering the walls is~\cite{Ellis:2019oqb,Lewicki:2022pdb}
\begin{equation}
K_w\simeq\max\left(0,1-\frac{P_{\rm LO}}{\Delta V_0}\right).
\label{reh:wall-efficiency}
\end{equation}
Here $P_{\rm LO}$ is the leading-order friction for ultrarelativistic walls with predominantly transmitted incoming particles~\cite{Bodeker:2009qy,Bodeker:2017cim}.
We evaluate it with Bose--Einstein scalar distributions, fermion chemical potentials $\pm\mu_\chi$ and curvature masses rather than hard thermal pole masses, and hold $K_w$ fixed.
This estimate does not determine particle transport or the gradient energy at collision, which can also depend on the finite acceleration time~\cite{Ellis:2019oqb}.

Using the envelope removal of collided segments~\cite{Kosowsky:1992vn} and the survival probability of uncollided wall elements~\cite{Lewicki:2022pdb}, we model wall-energy loss per unit total volume and unit $N=\ln(a/a_p)$ as
\begin{equation}
\mathcal D_w=-\rho_w\frac{\dd\ln P_{\rm false}}{\dd N}=I'\rho_w,
\label{reh:wall-energy-transfer}
\end{equation}
where a prime denotes $\dd/\dd N$, $P_{\rm false}=e^{-I}$, and $I$ follows from Eq.~\eqref{pt:percolation}.
We assume that dissipated wall energy thermalizes promptly within the broken-phase dark bath.
Scalar decays and inverse decays then transfer energy between this bath and the visible plasma at the finite rate $\mathcal C_{E,\mathrm{true}}$.

We define the broken-phase energy per unit total physical volume as $\bar\rho_{\rm true}=P_{\rm true}\rho_{\rm true}$.
The total energy density and Hubble rate are
\begin{equation}
\rho_{\rm tot}=P_{\rm false}\rho_{\rm false}+\bar\rho_{\rm true}+\rho_w+\rho_{\rm SM},\qquad H^2=\frac{\rho_{\rm tot}}{3\Mpl^2}.
\label{reh:total-energy}
\end{equation}
The coupled energy equations, including expansion, phase conversion and heat transfer, are
\begin{equation}
\begin{split}
&\rho_{\mathrm{false}}'+3(\rho_{\mathrm{false}}+p_{\mathrm{false}})={}-\mathcal C_{E,\mathrm{false}}/H,\\
&\bar\rho_{\mathrm{true}}'+3(\bar\rho_{\mathrm{true}}+P_{\mathrm{true}}p_{\mathrm{true}})={}(\rho_{\mathrm{false}}-K_w\Delta V_0)P_{\mathrm{true}}'+\mathcal D_w-P_{\mathrm{true}}\mathcal C_{E,\mathrm{true}}/H,\\
&\rho_w'+4\rho_w={}K_w\Delta V_0P_{\mathrm{true}}'-\mathcal D_w,\\
&\rho_{\rm SM}'+3(\rho_{\rm SM}+p_{\rm SM})={}(P_{\mathrm{false}}\mathcal C_{E,\mathrm{false}}+P_{\mathrm{true}}\mathcal C_{E,\mathrm{true}})/H.
\end{split}
\label{reh:phasewise-energy}
\end{equation}
Equation~\eqref{pt:phase-thermodynamics} determines the phase temperatures and chemical potentials from $d$ and the local densities, with $\rho_{\rm true}=\bar\rho_{\rm true}/P_{\rm true}$ for $P_{\rm true}>0$; the visible equation of state determines $T_\gamma$.
The transfer terms conserve total energy, including the charge-carrier rest energy already contained in the phase densities.
At $P_{\rm false}=10^{-8}$, we redistribute the remaining symmetric-phase energy with the same fractions, without adding vacuum energy again, and continue wall dissipation on the scale set by the bubble geometry.

The late homogeneous bath has $\rho_D=\rho_B+n\overline E_\chi$ and $p_D=p_B+nT_D$, where $\rho_B,p_B$ use scalar vacuum masses and Bose--Einstein distributions at zero chemical potential.
The fermions share the temperature $T_D$ and obey Maxwell--Boltzmann statistics.
We evolve their total number density $n=n_\chi+n_{\overline\chi}$ independently.
Their mean energy at $T_D$ is $\overline E_\chi=M_\chi K_1(M_\chi/T_D)/K_2(M_\chi/T_D)+3T_D$~\cite{EscuderoAbenza:2020cmq}.
After completion and wall dissipation, we match the cooling broken phase to this bath at time $t_h$, when $M_\chi/T_{\rm true}=16$, preserving the scale factor, photon temperature, charge and total dark energy.
\begin{equation}
\rho_{D,h}=\rho_{\mathrm{true}}(t_h^-)+\rho_w(t_h^-),\qquad a_h=a(t_h),\qquad a_p=1.
\label{reh:initial-budget}
\end{equation}
With $n_h$ the equilibrium fermion density at fixed charge from Section~\ref{sec:charge-equilibrium}, the initial temperature satisfies
\begin{equation}
\rho_B(T_{D,h})+n_h\overline E_\chi(T_{D,h})=\rho_{D,h}.
\label{reh:initial-temperature}
\end{equation}
The phase-transition and homogeneous-bath stages use heavy-scalar vacuum masses $457.2\MeV$ and $453.5\MeV$ for B1, respectively; matching conserves total energy and charge, not individual scalar populations.
Post-collision scalar oscillations and finite damping times~\cite{Cutting:2018tjt,Cutting:2020nla,Jinno:2017fby} can delay thermalization within the dark bath~\cite{Berges:2002wr,Micha:2004bv}.

In the homogeneous bath after the transition, the dark and visible energy densities evolve according to~\cite{EscuderoAbenza:2020cmq}
\begin{equation}
\begin{split}
\dot\rho_D+3H(\rho_D+p_D)={}&-\mathcal C_E,\\
\dot\rho_{\rm SM}+3H(\rho_{\rm SM}+p_{\rm SM})={}&\mathcal C_E,\qquad H^2=\frac{\rho_D+\rho_{\rm SM}}{3\Mpl^2}.
\end{split}
\label{reh:energy-equations}
\end{equation}
Writing $\dot n+3Hn=C_n$ gives
\begin{equation}
\left(\frac{\dd\rho_B}{\dd T_D}+n\frac{\dd\overline E_\chi}{\dd T_D}\right)\dot T_D=-3H(\rho_B+p_B+nT_D)-\mathcal C_E-\overline E_\chi C_n.
\label{reh:temperature-equation-main}
\end{equation}
The term $-\overline E_\chi C_n$ accounts for annihilation heating of the remaining dark bath.

The visible equation of state uses Ref.~\cite{Saikawa:2018rcs} in the QCD temperature range and a lepton--photon gas at low temperatures, with pressure obtained from $\dd(p/T)/\dd\ln T=\rho/T$.
The comoving entropy of the heated bath obeys $\dd(a^3s)/\dd t=a^3\mathcal C_E/T_\gamma$~\cite{Ertas:2021xeh}.
The primary histories use a common visible temperature and assume re-equilibration before three massless neutrino species decouple instantaneously at $T_\gamma=3\MeV$ on the final cooling branch, at time $t_{\rm sep}$.
Thereafter, $T_\nu\propto a^{-1}$ and heat enters only the electromagnetic plasma, whose comoving entropy $S_{\rm EM}=a^3s_{\rm EM}$ reaches
\begin{equation}
S_{\mathrm{EM},\mathrm{final}}=S_{\rm EM}(t_{\rm sep})+\int_{t_{\rm sep}}^{t_{\mathrm{final}}}\dd t\,\frac{a^3\mathcal C_E}{T_\gamma}.
\label{reh:integrated-em-entropy}
\end{equation}
During conversion, $\mathcal C_E$ denotes the volume-weighted sum of the energy transfer rates in the two phases.
We integrate until heat transfer is negligible; subsequent adiabatic $e^+e^-$ annihilation preserves this entropy even before the temperature ratio reaches its asymptotic value.
The BBN calculations in Section~\ref{sec:bbn-delayed-heating} use continuous neutrino energy exchange for the AD inputs of each benchmark, with a thermal treatment of neutrino energies that omits spectral distortions and flavour oscillations~\cite{Mangano:2005cc,Froustey:2020mcq}.
Bounds on late-decaying matter also depend on the heating history~\cite{Kawasaki:2000en,deSalas:2015glj}.

\begin{figure}[!htbp]
\centering
\includegraphics[width=0.45\textwidth]{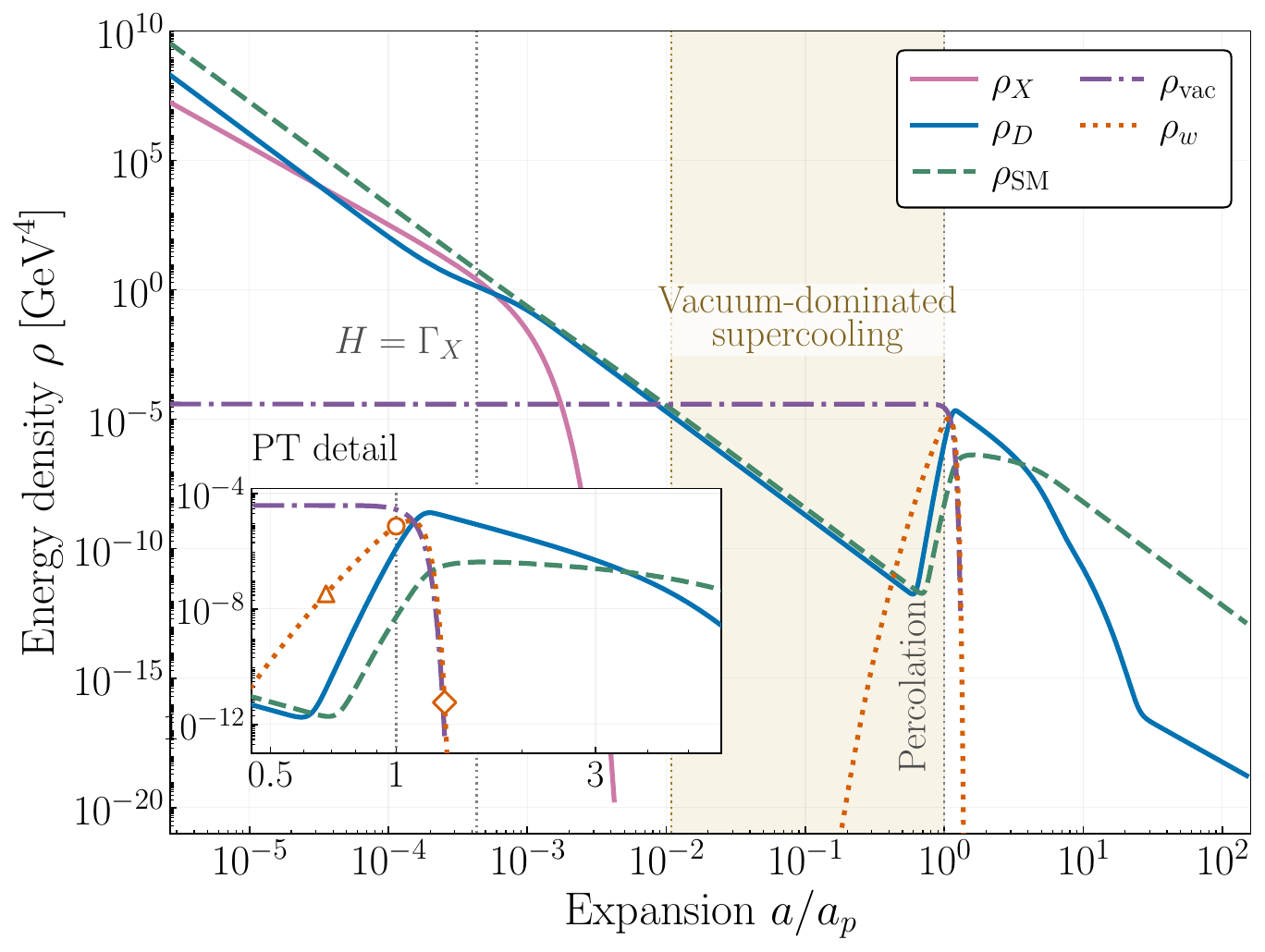}\hfill
\includegraphics[width=0.45\textwidth]{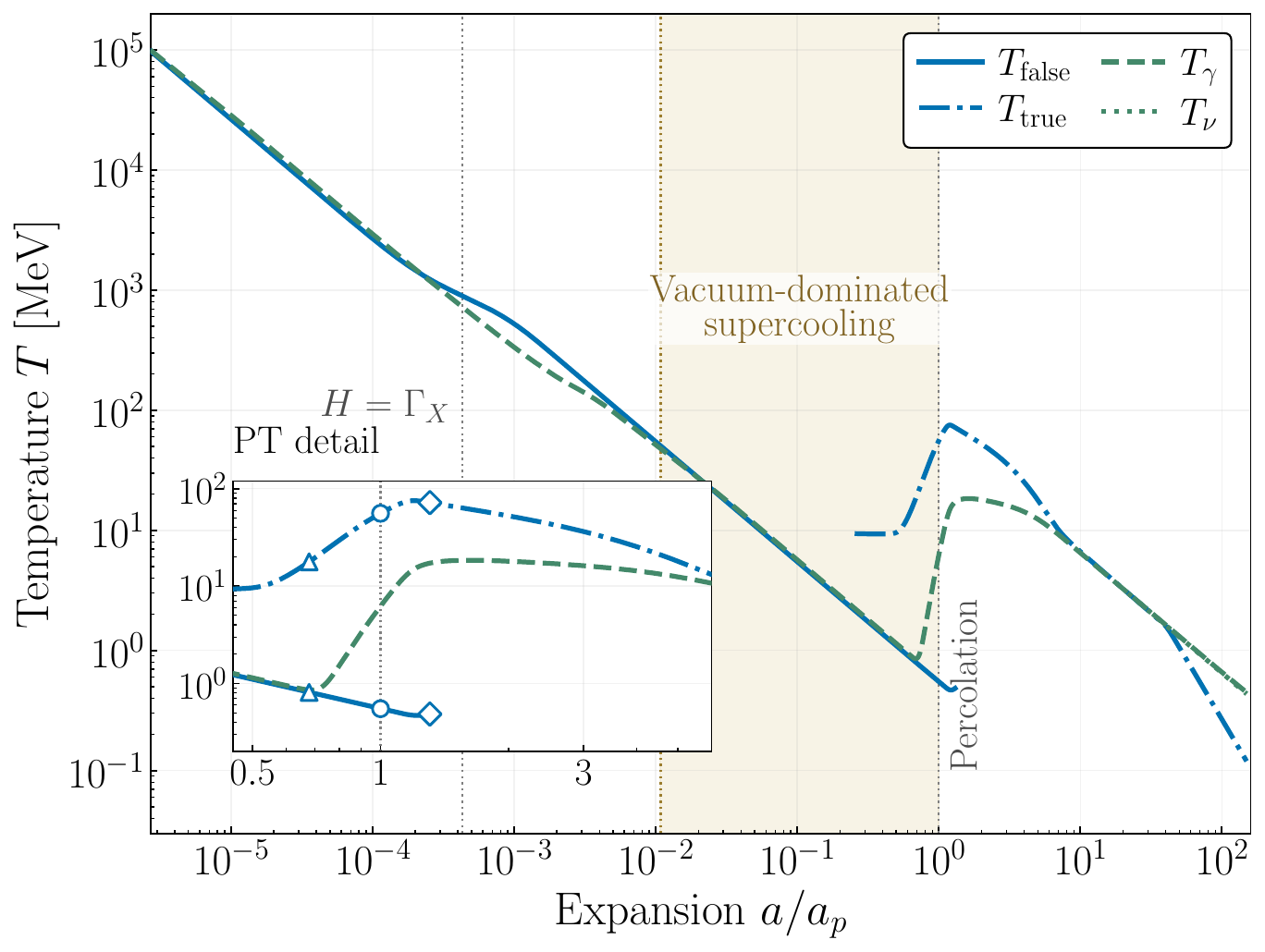}
\caption{Energy densities (left) and temperatures (right) for B1 during AD decay, phase transition and late reheating.
The left panel shows the oscillation-averaged condensate density $\rho_X$, the dark-bath density $\rho_D$, the visible-bath density $\rho_{\rm SM}$ including neutrinos, the remaining vacuum energy density $\rho_{\rm vac}$, and the bubble-wall energy density $\rho_w$ (orange dotted curve).
During phase conversion, $\rho_D=P_{\rm false}(\rho_{\rm false}-\Delta V_0)+P_{\rm true}\rho_{\rm true}$ and $\rho_{\rm vac}=P_{\rm false}\Delta V_0$.
The right panel shows $T_{\rm false}$ for the symmetric-phase dark sector, and $T_{\rm true}$ for the broken-phase dark sector.
The triangle, circle and diamond mark unit nucleation ($\mathcal N_{\rm nuc}=1$), percolation and numerical completion at $P_{\mathrm{false}}=10^{-8}$, respectively.
These markers appear on $\rho_w$ in the left inset and on both $T_{\rm false}$ and $T_{\rm true}$ in the right inset.}
\label{fig:hybrid-reheating}
\end{figure}

Figure~\ref{fig:hybrid-reheating} connects the AD evolution, phase transition and reheating stages of B1.
The broken-phase temperature reaches $75.48\MeV$ at $a/a_p=1.198$; the photon temperature peaks later at $T_{\gamma,\max}=18.42\MeV$ and $a/a_p=1.623$, where the maximum refers to phase-transition reheating and excludes the earlier AD epoch.
Entropy production continues beyond this peak, and the dark energy density eventually approaches $M_\chi Q_\chi a^{-3}$.

\FloatBarrier

\section{Gravitational waves}
\label{sec:gw}

The released vacuum energy can enter scalar gradients or fluid motion~\cite{Kamionkowski:1993fg,Espinosa:2010hh,Ellis:2019oqb}.
For the collision spectrum, we identify $\kappa_{\rm coll}=K_w$, with $K_w$ calculated using Eq.~\eqref{reh:wall-efficiency}.
To normalize the collision spectrum, we define $\alpha_p$ as the ratio of released vacuum energy to incoming non-vacuum energy at percolation and $K_\phi$ as the energy fraction entering the collision model~\cite{Matuszak:2026xsz}:
\begin{equation}
\begin{split}
K_\phi&=K_w\frac{\alpha_p}{1+\alpha_p}=\frac{K_w\Delta V_0}{\rho_{\mathrm{false}}(T_{\mathrm{false},p},\eta_{\chi,\mathrm{false},p})+\rho_{{\rm SM},p}},\\
\alpha_p&=\frac{\Delta V_0}{\rho_{\mathrm{false}}(T_{\mathrm{false},p},\eta_{\chi,\mathrm{false},p})-\Delta V_0+\rho_{{\rm SM},p}},
\end{split}
\label{gw:lo_fraction}
\end{equation}
where the denominator of $K_\phi$ contains the incoming energy densities, while $H_*\equiv H(t_p)$ is the Hubble rate at percolation, calculated from the total energy density in Eq.~\eqref{reh:total-energy}.
Thus $K_w$ is a fraction of the released vacuum energy, $K_\phi$ is the fraction used in the spectrum, and $\rho_w/\rho_{\rm tot}$ is the instantaneous wall fraction.

Motivated by the leading-order friction estimate, we assume that bubble collisions dominate the GW signal and use the broken power-law fit for a real scalar field from Ref.~\cite{Lewicki:2020jiv}.
We evaluate the emission parameters at percolation, with no additional converted-volume factor, and write the fit in terms of the mean bubble separation $R_*$ and the conversion factor $C_R$ in Eq.~\eqref{pt:beta_separation}.
For ordinary frequency $f_*$ at emission, $\Omega_{\rm GW,*}=\Omega_{\rm col,*}$ denotes the GW energy density per logarithmic frequency interval, normalized to $3\Mpl^2H_*^2$.
With spectral shape $\mathcal S$, overall normalization $A_{\rm col}$ and characteristic frequency $f_{c,*}$, the collision spectrum is
\begin{equation}
\Omega_{\mathrm{col},*}(f_*)=\frac{A_{\mathrm{col}}}{C_R^2}(H_*R_*)^2K_\phi^2\mathcal S(f_*),\qquad A_{\mathrm{col}}=0.0471,\qquad f_{c,*}=\frac{0.83C_R}{2\pi R_*}.
\label{gw:source}
\end{equation}
The fitted spectral shape is
\begin{equation}
\mathcal S(f_*)=\left[\frac{a+b}{b x^{-a/c}+a x^{b/c}}\right]^c,\qquad x=\frac{f_*}{f_{c,*}},\qquad (a,b,c)=(2.84,2.29,2.52).
\label{gw:source-shape}
\end{equation}
This shape peaks at $f_*=f_{c,*}$ with $\mathcal S(f_{c,*})=1$.

The present-day GW spectrum is given by~\cite{Ertas:2021xeh}
\begin{equation}
f_0=\frac{a_p}{a_0}f_*,\qquad h^2\Omega_{\mathrm{GW},0}(f_0)=\left(\frac{H_*}{H_{100}}\right)^2\left(\frac{a_p}{a_0}\right)^4\Omega_{\mathrm{GW},*}(f_*),
\label{gw:redshift}
\end{equation}
where $a_p=a(t_p)$ is the scale factor at percolation, the subscript $0$ denotes present-day quantities, $H_{100}=100\,\mathrm{km\,s^{-1}\,Mpc^{-1}}$, and $h=H_0/H_{100}$.
With $a_p=1$, the final electromagnetic entropy in Eq.~\eqref{reh:integrated-em-entropy} determines the redshift factor:
\begin{equation}
\frac{a_p}{a_0}=\left(\frac{s_{\gamma0}}{S_{\mathrm{EM},\mathrm{final}}}\right)^{1/3},\qquad s_{\gamma0}=\frac{4\pi^2}{45}T_{\gamma0}^3,
\label{reh:late-redshift}
\end{equation}
where $s_{\gamma0}$ and $T_{\gamma0}$ are the present-day photon entropy density and temperature, respectively.
The redshift factor obtained from the calculated $S_{\mathrm{EM},\mathrm{final}}$ accounts for expansion through conversion and reheating without imposing the standard neutrino-to-photon ratio.

Combining Eqs.~\eqref{gw:source}, \eqref{gw:redshift} and \eqref{reh:late-redshift} gives the redshifted characteristic frequency $f_{c,0}$ and the amplitude evaluated at that frequency:
\begin{equation}
\begin{split}
f_{c,0}={}&\frac{0.83C_R}{2\pi}\frac{H_*}{H_*R_*}\left(\frac{s_{\gamma0}}{S_{\mathrm{EM},\mathrm{final}}}\right)^{1/3},\\
h^2\Omega_{\rm GW,0}(f_{c,0})={}&\frac{A_{\rm col}}{C_R^2}K_\phi^2(H_*R_*)^2\left(\frac{H_*}{H_{100}}\right)^2\left(\frac{s_{\gamma0}}{S_{\mathrm{EM},\mathrm{final}}}\right)^{4/3}.
\end{split}
\label{gw:source-entropy-factorization}
\end{equation}
The powers of $S_{\mathrm{EM},\mathrm{final}}$ in Eq.~\eqref{gw:source-entropy-factorization} show that entropy production lowers both frequency and amplitude at fixed emission parameters~\cite{Ertas:2021xeh}.

\begin{figure}[!htbp]
\centering
\includegraphics[width=0.75\textwidth]{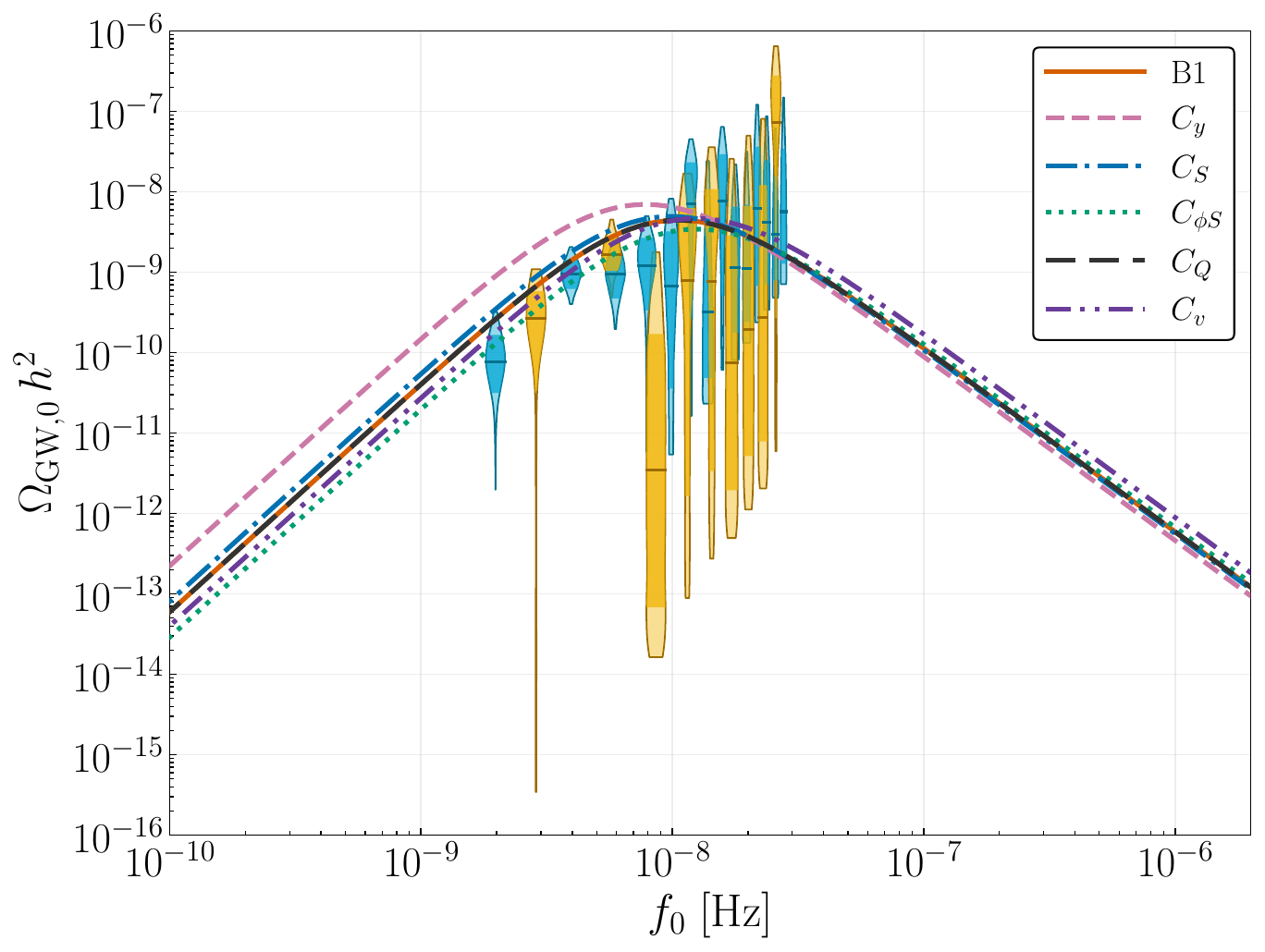}
\caption{Gravitational-wave spectra from bubble collisions for various benchmarks in Table~\ref{tab:benchmark-summary}, compared with NANOGrav CNM (cyan) and EPTA DR2new+ (amber) free-spectrum posteriors in individual frequency bins.}
\label{fig:early-hybrid-cnm}
\end{figure}

Figure~\ref{fig:early-hybrid-cnm} compares the spectra with the Hellings--Downs free-spectrum posteriors in individual frequency bins from the NANOGrav custom chromatic noise model (CNM)~\cite{Agarwal:2026dlv,Larsen:2026pru} and EPTA DR2new+~\cite{EPTA:2023fyk}.
For this visual comparison, we approximate the CNM posteriors from the first $14$ blue contours in Figure~1 of Ref.~\cite{Agarwal:2026dlv}, and obtain the EPTA posteriors from the released DR2new+ samples~\cite{EPTA:2023fyk}.

\section{Asymmetric dark matter}
\label{sec:asymdm}
\subsection{Conserved charge and kinetic assumptions}
\label{sec:charge-equilibrium}
The reactions $\chi\overline\chi\leftrightarrow\sigma\sigma,H_DH_D,A_SA_S$ preserve the fermion asymmetry; once AD charge transfer and washout are negligible, $d=Q_\chi a^{-3}$ with constant $Q_\chi$~\cite{Kaplan:2009ag,Graesser:2011wi,Zurek:2013wia}.
We assume a common kinetic temperature $T_D$ and zero scalar chemical potentials, so chemical equilibrium requires $\mu_{\overline\chi}^{\rm eq}=-\mu_\chi^{\rm eq}$~\cite{Iminniyaz:2011yp}.
The Maxwell--Boltzmann equilibrium densities are~\cite{EscuderoAbenza:2020cmq}
\begin{equation}
n_{\chi,{\rm eq},Q_\chi}=\frac{n_0}{2}e^{\mu_\chi^{\rm eq}/T_D},\qquad n_{\overline\chi,{\rm eq},Q_\chi}=\frac{n_0}{2}e^{-\mu_\chi^{\rm eq}/T_D},
\label{adm:finite-mu-densities}
\end{equation}
where $n_0(T_D)=2M_\chi^2T_DK_2(M_\chi/T_D)/\pi^2$ is the total equilibrium fermion density at zero chemical potential, and $K_\nu$ denotes a modified Bessel function of the second kind.
Fixing the density difference to $d$ determines the chemical potential and equilibrium densities~\cite{Iminniyaz:2011yp}:
\begin{equation}
\begin{split}
\frac{\mu_\chi^{\rm eq}}{T_D}={}&\operatorname{arsinh}\left(\frac{d}{n_0}\right),\qquad n_{{\rm eq},Q_\chi}=\sqrt{d^2+n_0^2},\\
n_{\overline\chi,{\rm eq},Q_\chi}={}&\frac{n_0^2}{2(\sqrt{d^2+n_0^2}+d)},\qquad n_{\chi,{\rm eq},Q_\chi}=n_{\overline\chi,{\rm eq},Q_\chi}+d.
\end{split}
\label{adm:equilibrium}
\end{equation}
The total density $n_h=n_{{\rm eq},Q_\chi}(T_{D,h},a_h)$ enters the energy matching in Eq.~\eqref{reh:initial-temperature}.
During chemical equilibrium, charge dilution gives $\left(\partial n_Q/\partial\ln a\right)_{T_D}=-3d^2/n_Q$, where $n_Q=n_{{\rm eq},Q_\chi}$.
Including this dependence in the energy equation gives the temperature evolution and heat-capacity coefficient $\mathcal C_Q$:
\begin{equation}
\begin{split}
\mathcal C_Q={}&\frac{\dd\rho_B}{\dd T_D}+n_Q\frac{\dd\overline E_\chi}{\dd T_D}+\overline E_\chi\frac{n_0}{n_Q}\frac{\dd n_0}{\dd T_D},\\
\mathcal C_Q\dot T_D={}&-3H\left(\rho_B+p_B+n_QT_D+\overline E_\chi\frac{n_0^2}{n_Q}\right)-\mathcal C_E.
\end{split}
\label{adm:equilibrium-temperature-main}
\end{equation}

The annihilation and elastic-scattering amplitudes use vacuum masses and zero-momentum vertices from $\Veff$ at $(v,0)$, including $g_{\sigma\sigma\sigma}=40Bv$ and $g_{\sigma ii}=\partial_\phi\partial_i^2\Veff$ for real scalars $i$, and neglect momentum-dependent loop corrections.
We estimate the momentum-relaxation rate $\Gamma_{\rm mom}$ from $\chi\sigma\to\chi\sigma$ with fermion and real-scalar exchange, averaging over initial spins and assuming small momentum transfer~\cite{Binder:2017rgn}.
On the selected histories, this rate falls below $H$ only after the antiparticle population becomes negligible.
Scalar chemical equilibrium and initial relativistic thermalization remain assumptions~\cite{Arcadi:2019oxh,Hufnagel:2022aiz}.
We do not evolve separate species temperatures or establish that chemical-relaxation eigenrates exceed the rates of expansion, visible-sector energy loss and other processes driving the bath out of equilibrium.

\subsection{Annihilation, entropy production and relic abundance}
\label{sec:benchmark-relic}
We evolve the antiparticle density in $N=\ln(a/a_p)$ using~\cite{Iminniyaz:2011yp}
\begin{equation}
\frac{\dd n_{\overline\chi}}{\dd N}=-3n_{\overline\chi}-\frac{\langle\sigma v\rangle}{H}\left[n_{\overline\chi}(n_{\overline\chi}+d)-\frac{n_0^2}{4}\right],\qquad n_\chi=n_{\overline\chi}+d.
\label{adm:minority}
\end{equation}
The Maxwell--Boltzmann thermal average sums the scalar final states~\cite{Gondolo:1990dk}:
\begin{equation}
\langle\sigma v\rangle=\sum_{i=\sigma,H_D,A_S}\frac{1}{8M_\chi^4T_DK_2^2(M_\chi/T_D)}\int_{4M_\chi^2}^{\infty}\dd s\,\sigma_{\chi\overline\chi\to ii}(s)(s-4M_\chi^2)\sqrt{s}\,K_1(\sqrt{s}/T_D).
\label{adm:annihilation-average-main}
\end{equation}
Here $s$ is the squared centre-of-mass energy and $\sigma_{\chi\overline\chi\to ii}$ is the cross section for each channel.
The thermal integral includes heavy final states inaccessible to fermions at rest~\cite{DAgnolo:2015ujb}.

The collision term for the total fermion density is twice the antiparticle collision term and couples to the temperature through Eq.~\eqref{reh:temperature-equation-main}.
We use Eq.~\eqref{adm:equilibrium-temperature-main} while pair relaxation is rapid, then switch to the coupled number and temperature equations before antiparticles become negligible.
Omitting annihilation loss after a later epoch $N_b$ while including inverse production gives an upper bound on the antiparticle fraction at the final epoch $N_{\mathrm{final}}$:
\begin{equation}
\frac{n_{\overline\chi}(N_{\mathrm{final}})}{d(N_{\mathrm{final}})}\leq\frac{n_{\overline\chi}(N_b)}{d(N_b)}+\int_{N_b}^{N_{\mathrm{final}}}\dd N\,\frac{a^3\langle\sigma v\rangle n_0^2}{4H Q_\chi}.
\label{adm:minority-bound-main}
\end{equation}

\begin{figure}[!tbp]
\centering
\includegraphics[width=0.75\textwidth]{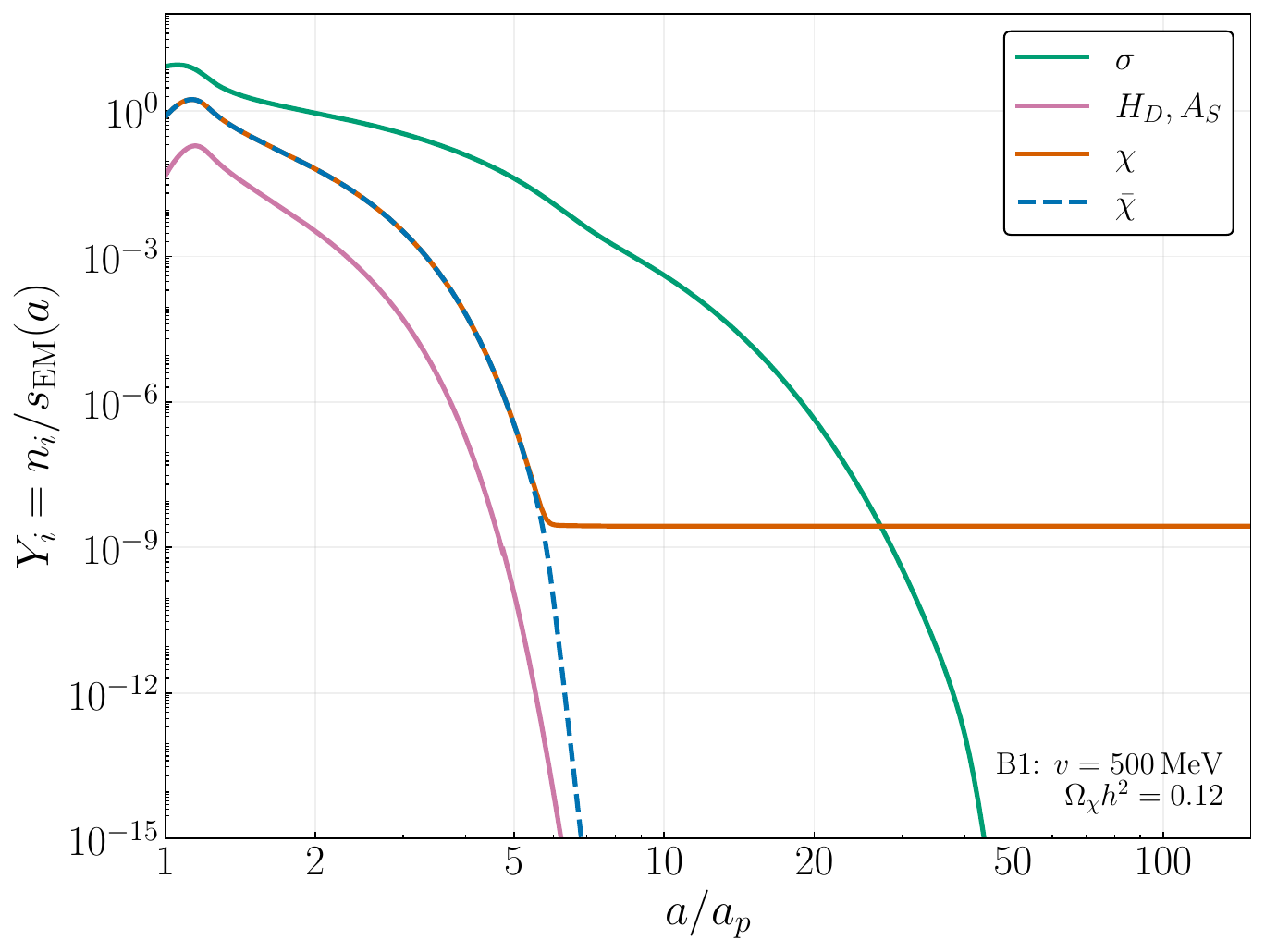}
\caption{Dark sector particle abundances for B1, with $M_\chi=312.5~\MeV$.
During coexistence, $Y_i=P_{\rm true}n_{i,\rm true}/s_{\rm EM}$ excludes the contribution from the false vacuum.
The $H_D,A_S$ curve represents each species separately; antiparticles are shown only while numerically resolved.}
\label{fig:dark-particle-abundances}
\end{figure}

After electromagnetic entropy production ends, the relic abundance follows from the conserved charge and residual fraction $r_{\overline\chi}=n_{\overline\chi}/d$, where the present critical density is denoted by $\rho_{c,0}$:
\begin{equation}
\begin{split}
\Omega_\chi h^2&=\frac{M_\chi s_{\gamma0}}{\rho_{c,0}/h^2}\frac{Q_\chi}{S_{\mathrm{EM},\mathrm{final}}}(1+2r_{\overline\chi})\\
&\simeq0.12\,\bfrac{M_\chi}{312.5\MeV}\bfrac{Q_\chi/S_{\mathrm{EM},\mathrm{final}}}{2.737\times10^{-9}}(1+2r_{\overline\chi}).
\end{split}
\label{adm:abundance}
\end{equation}
The numerical normalization is for B1 with instantaneous neutrino decoupling.
Figure~\ref{fig:dark-particle-abundances} shows its particle yields, and Eq.~\eqref{adm:minority-bound-main} gives $n_{\overline\chi}/d<1.586\times10^{-15}$ if no later particle production or reheating occurs.

The same final entropy controls relic dilution and GW redshift through Eqs.~\eqref{adm:abundance} and \eqref{gw:source-entropy-factorization}.
When antiparticles are negligible in both benchmarks, the abundance relative to B1 is
\begin{equation}
\frac{\Omega_\chi}{\Omega_\chi(\mathrm{B1})}=\frac{M_\chi}{M_\chi(\mathrm{B1})}\frac{Q_\chi}{Q_\chi(\mathrm{B1})}\frac{S_{\mathrm{EM},\mathrm{final}}(\mathrm{B1})}{S_{\mathrm{EM},\mathrm{final}}}.
\label{adm:source-response-factorization}
\end{equation}
With $a_p=1$ for each benchmark, $Q_\chi$ is the asymmetry density at percolation and includes dilution since AD charge transfer; the ratio $Q_\chi/S_{\mathrm{EM},\mathrm{final}}$ is independent of this normalization.
Section~\ref{sec:controlled-response} uses Eq.~\eqref{adm:source-response-factorization} to interpret the relic's parameter dependence.

\FloatBarrier

\section{Benchmark results and parameter dependence}
\label{sec:benchmark-results}
\subsection{Benchmark definitions and predictions}
\label{sec:benchmark-inputs}
Table~\ref{tab:benchmark-summary} collects the model inputs, transition and reheating quantities, relic abundances and GW peaks for the benchmark parameter sets, together with the joint BBN tests for B1, $C_v$, $C_y$, $C_{\phi S}$, $C_S$, $C_Q$, $E_1$ and $M_4$.
For B1, $r_X$ is chosen to match the observed dark-matter abundance by recomputing the AD evolution, phase conversion and entropy production at each trial.
The benchmark histories follow visible-sector heating during and after phase conversion.
For the tabulated transition, reheating, relic and GW predictions, we approximate neutrino decoupling as instantaneous on the final cooling branch.
The dilution factor $\Delta_{\rm vis}$ in Eq.~\eqref{bbn:baryon-dilution} compares the comoving visible-sector entropy after reheating with its value before phase conversion.

For each dark-sector benchmark, we vary one input and hold the remaining inputs fixed at their B1 values.
The benchmark $C_v$ changes $v$ to $600\MeV$ at fixed dimensionless couplings, AD parameters and reheating interaction scales, while $C_Q$ changes only $r_X$, which controls the AD-generated dark charge.
The coupling variations are
\begin{equation}
C_y:\ y_\chi\to1.01y_\chi,\qquad C_S:\ \lambda_S\to1.10\lambda_S,\qquad C_{\phi S}:\ \lambda_{\phi S}\to1.01\lambda_{\phi S}.
\label{result:control-definitions}
\end{equation}
The masses, matching and thermal histories are recalculated after each change, with $\lambda_\phi(\mu_{\rm GW})=0$ and the common inputs specified in Section~\ref{sec:asymmetry-origin}.
The reheating benchmarks $E_1$, $E_4$ and $M_4$ instead vary one lepton coupling while keeping B1's initial phase energy, charge and wall-energy fraction unchanged, without recomputing the earlier AD evolution.
The B1 couplings are $g_{e,0}=6.362\times10^{-10}$ and $g_{\mu,0}=3.030\times10^{-10}$.

\begin{table}[!tbp]
\centering
\footnotesize
\setlength{\tabcolsep}{2.4pt}
\renewcommand{\arraystretch}{1.22}
\begin{tabular}{@{}p{98pt}*{9}{>{\raggedleft\arraybackslash}p{33.2pt}}@{}}
\toprule
& \multicolumn{6}{c}{Dark-sector benchmarks} & \multicolumn{3}{c}{Reheating benchmarks} \\
\cmidrule(lr){2-7}\cmidrule(l){8-10}
Quantity & B1 & $C_v$ & $C_y$ & $C_{\phi S}$ & $C_S$ & $C_Q$ & $E_1$ & $E_4$ & $M_4$ \\
\midrule
\multicolumn{10}{@{}l}{\itshape Model inputs} \\
$v\,[\MeV]$ & 500 & 600 & 500 & 500 & 500 & 500 & 500 & 500 & 500 \\
$y_\chi$ & 0.6251 & 0.6251 & 0.6313 & 0.6251 & 0.6251 & 0.6251 & 0.6251 & 0.6251 & 0.6251 \\
$\lambda_{\phi S}$ & 1.672 & 1.672 & 1.672 & 1.689 & 1.672 & 1.672 & 1.672 & 1.672 & 1.672 \\
$\lambda_S$ & 0.2428 & 0.2428 & 0.2428 & 0.2428 & 0.267 & 0.2428 & 0.2428 & 0.2428 & 0.2428 \\
$10^3r_X$ & 37.94 & 37.94 & 37.94 & 37.94 & 37.94 & 43.22 & 37.94 & 37.94 & 37.94 \\
$g_e/g_{e,0}$ & 1 & 1.2 & 1 & 1 & 1 & 1 & 10 & $10^4$ & 1 \\
$g_\mu/g_{\mu,0}$ & 1 & 1.2 & 1 & 1 & 1 & 1 & 1 & 1 & $10^4$ \\
\midrule
\multicolumn{10}{@{}l}{\itshape Phase transition and reheating} \\
$T_{\mathrm{false},p}\,[\MeV]$ & 0.5499 & 0.6248 & 0.2697 & 0.8258 & 0.4846 & 0.5498 & 1.001 & 17.19 & 16.18 \\
$T_{\mathrm{false},\mathrm{comp}}\,[\MeV]$ & 0.4848 & 0.6095 & 0.462 & 0.6715 & 0.46 & 0.4847 & 6.846 & --- & 32.87 \\
$T_{\gamma,\max}\,[\MeV]$ & 18.42 & 23.02 & 18.15 & 18.64 & 18.39 & 18.42 & 38.82 & 20.56 & 41.02 \\
$\Delta_{\rm vis}/10^6$ & 1.515 & 1.781 & 13.26 & 0.4499 & 2.240 & 1.516 & 1.441 & --- & 1.551 \\
$H_*R_*$ & 0.3075 & 0.3151 & 0.4031 & 0.2659 & 0.3275 & 0.3076 & 0.4369 & --- & 0.5504 \\
Complete & Yes & Yes & Yes & Yes & Yes & Yes & Yes & No & Yes \\
\midrule
\multicolumn{10}{@{}l}{\itshape Relic abundance and GW peaks} \\
$\Omega_\chi h^2$ & 0.12 & 0.1225 & 0.01385 & 0.4041 & 0.08116 & 0.1365 & 0.1262 & --- & 0.1172 \\
$f_{\rm peak,0}\,[\mathrm{nHz}]$ & 10.61 & 12.43 & 7.817 & 12.52 & 9.906 & 10.61 & 7.612 & --- & 5.923 \\
$h^2\Omega_{\rm GW,peak,0}/10^{-9}$ & 4.447 & 4.673 & 6.977 & 3.445 & 4.948 & 4.448 & 9.529 & --- & 13.45 \\
\midrule
\multicolumn{10}{@{}l}{\itshape BBN test} \\
PArthENoPE & 1.116 & 1.256 & 3.478 & 0.6470 & 1.447 & 1.117 & 2.742 & --- & 4.875 \\
PRIMAT & 0.1203 & 0.09323 & 0.5633 & 0.3549 & 0.07685 & 0.1201 & 0.2919 & --- & 1.189 \\
\bottomrule
\end{tabular}
\caption{The four blocks list the model inputs, transition and reheating quantities, relic abundance and gravitational-wave peaks, and $\chi^2_{\rm BBN}$ for the joint D/H and $Y_{\rm P}$ test.}
\label{tab:benchmark-summary}
\end{table}

All tabulated histories except $E_4$ complete the transition within the evolved interval.
Visible-sector heating can raise the temperature of the dark sector in the false vacuum, so $T_{\mathrm{false},\mathrm{comp}}$ need not be below $T_{\mathrm{false},p}$.
The bubble separation and final electromagnetic entropy determine the GW peaks through Eq.~\eqref{gw:source-entropy-factorization}.
Figure~\ref{fig:early-hybrid-cnm} compares the resulting spectra with PTA free-spectrum posteriors; this is a comparison of spectral shapes and amplitudes, not a benchmark likelihood fit.

At fixed B1 charge and GW emission parameters, continuous neutrino energy exchange lowers the final electromagnetic entropy by $0.47\%$ relative to instantaneous neutrino decoupling.
The relic abundance then rises to $\Omega_\chi h^2=0.1206$, while the GW peak frequency and amplitude each shift by less than $1\%$.
For this homogeneous B1 history the neutrino energy density after electron--positron annihilation corresponds to $N_{{\rm eff},\nu}=3.044$.
Integrating the fitted collision spectrum over frequency, including its extrapolation beyond the PTA band, gives a GW contribution $\Delta N_{{\rm eff},{\rm GW}}=0.001371$~\cite{Maggiore:1999vm}.

\subsection{Dependence on dark-sector and reheating parameters}
\label{sec:controlled-response}
At fixed $v$, the leading vacuum response follows by differentiating the Coleman--Weinberg coefficient in Eq.~\eqref{mod:cw-coefficient}:
\begin{equation}
\delta\ln B=\frac{2\lambda_{\phi S}^2\delta\ln\lambda_{\phi S}-32y_\chi^4\delta\ln y_\chi}{\lambda_{\phi S}^2-8y_\chi^4}.
\label{result:controlled-vacuum-response}
\end{equation}
The same fractional response applies to $\Delta V_0$ and $m_{\sigma,\rm LO}^2$.
Thus $C_y$ lowers $B$, $C_{\phi S}$ raises it, and $\lambda_S$ affects screening and matching without entering $B$ at this order.
The transition histories also depend on the thermal terms and fluctuation corrections in Eq.~\eqref{pt:nlodet-rate}.
Because the coupling increments differ, the tabulated changes should not be compared directly as logarithmic sensitivities.

Charge conservation relates these changes in transition timing to the relic abundance.
Let $N_{\rm ph\to p}=\ln(a_p/a_{\rm ph,init})$ denote the expansion between phase initialization and percolation, and let $d_{\rm ph,init}$ be the initial charge density defined in Section~\ref{sec:source-phase-matching}.
For negligible antiparticle abundance, Eq.~\eqref{adm:source-response-factorization} gives
\begin{equation}
\Delta\ln(\Omega_\chi h^2)=\Delta\ln M_\chi+\Delta\ln d_{\mathrm{ph,init}}-3\Delta N_{\mathrm{ph}\to p}-\Delta\ln S_{\mathrm{EM},\mathrm{final}}.
\label{result:controlled-relic-response}
\end{equation}
Here all differences are relative to B1, with $a_p=1$ for each history and the final entropy evaluated after visible-sector energy transfer becomes negligible.
For $C_y$, $C_{\phi S}$ and $C_S$, the initial charge density changes negligibly and the final entropy changes modestly, so expansion before conversion accounts for most of the change in the relic abundance.
The later transitions in $C_y$ and $C_S$ dilute the conserved asymmetry more strongly, whereas the earlier transition in $C_{\phi S}$ leaves a larger relic abundance.
In $C_y$, this dilution outweighs the small increase in $M_\chi$.
The GW peaks vary less strongly, following the bubble-separation and entropy dependence in Eq.~\eqref{gw:source-entropy-factorization}.

The benchmark $C_Q$ increases the relic abundance with almost no shift in the GW peak, while $C_v$ mainly shifts the peak frequency and leaves the relic abundance close to its B1 value.
The two nearby values of $r_X$ establish the local response to this parameter.
Whether the relic abundance and the GW signal can vary independently over a wider range remains open.

\begin{figure}[!htbp]
\centering
\includegraphics[width=0.45\textwidth]{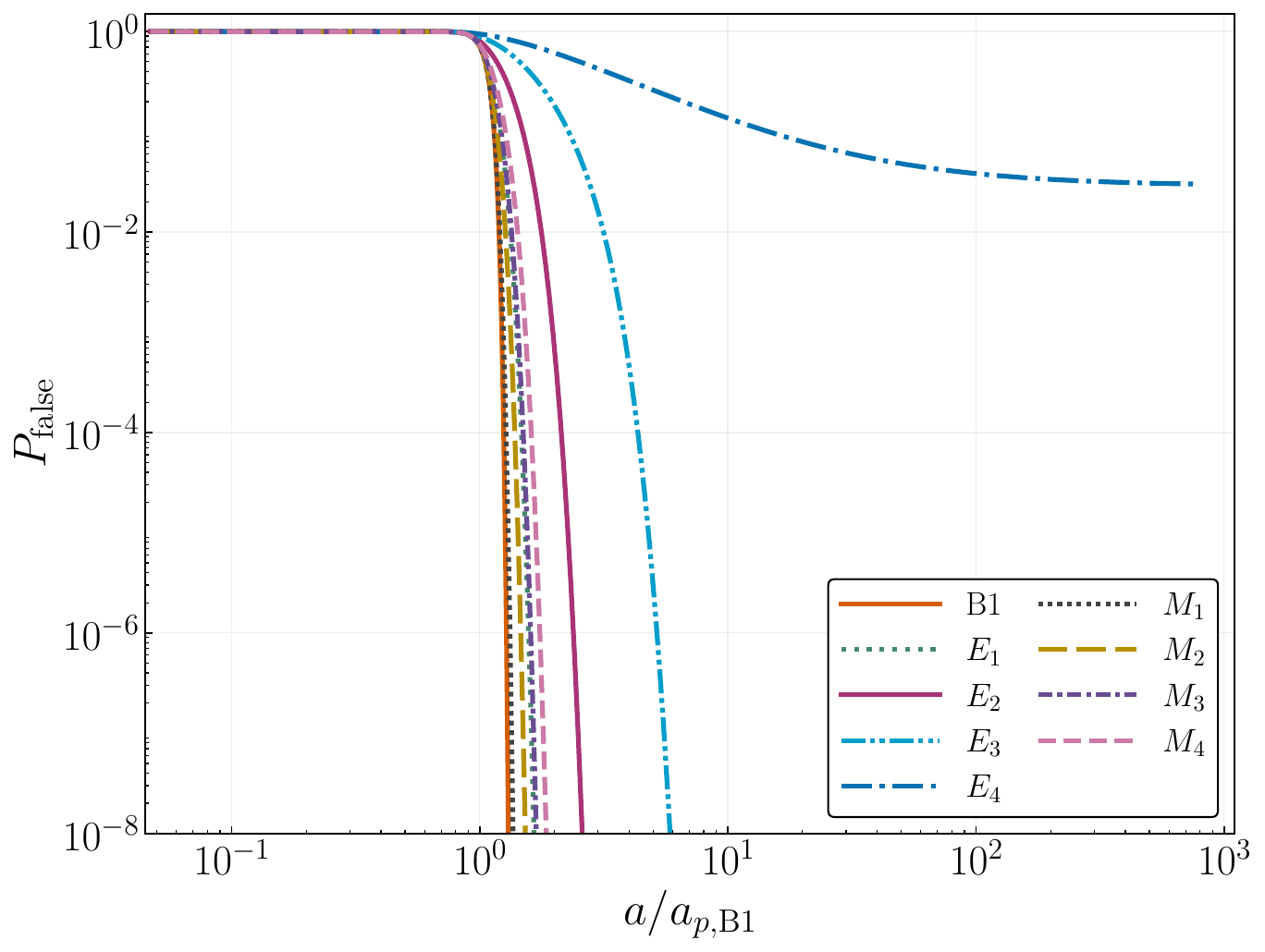}\hfill
\includegraphics[width=0.45\textwidth]{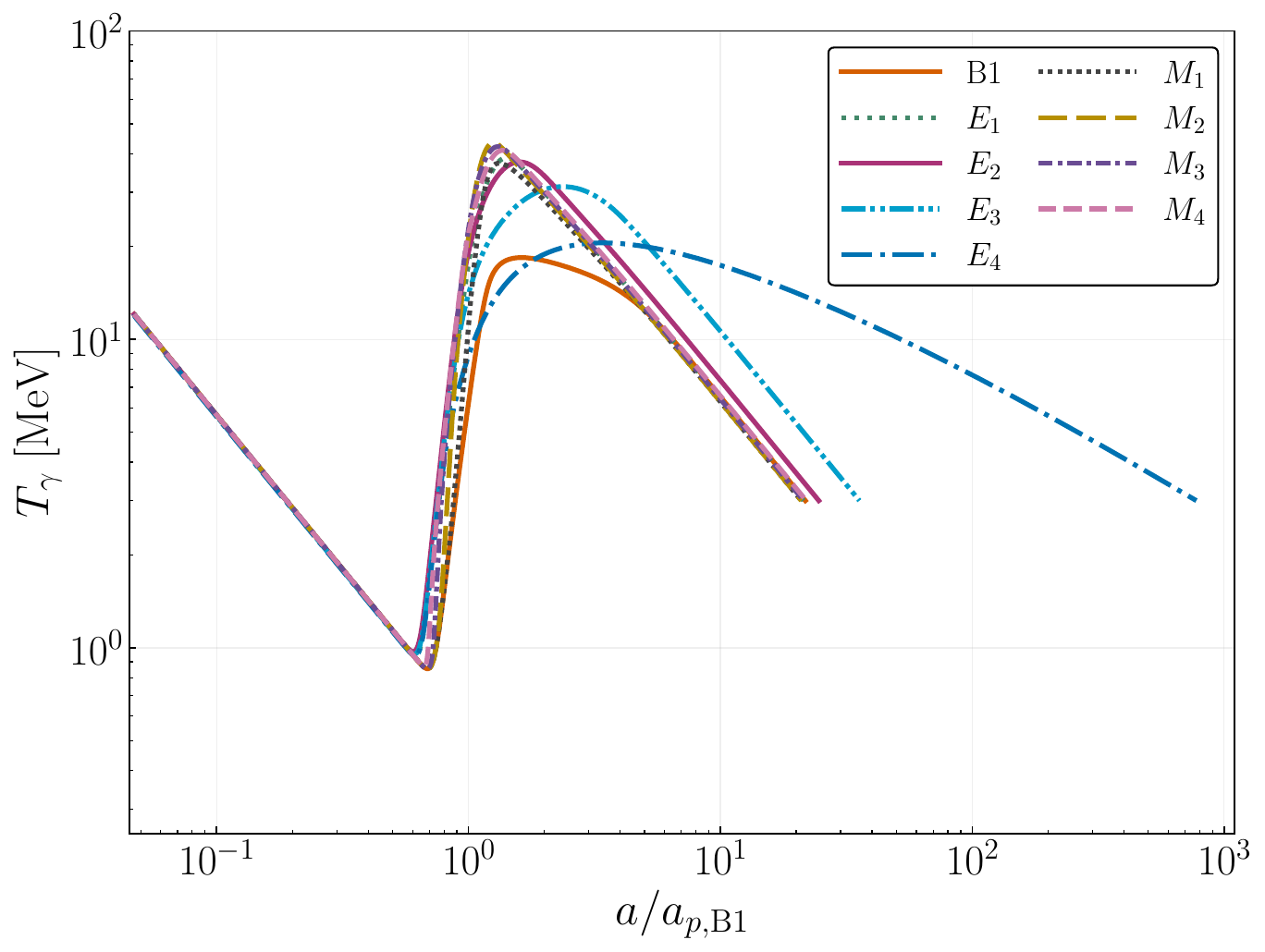}
\caption{False-vacuum fractions (left) and visible temperatures (right) for B1 and the reheating benchmarks $E_1$--$E_4$ and $M_1$--$M_4$.
The $E_i$ and $M_i$ benchmarks increase the electron and muon couplings, respectively, by a factor of $10^i$, with all other inputs held fixed at their B1 values.
The scale factor is given relative to its value at percolation for B1.}
\label{fig:visible-portal-controls}
\end{figure}

The reheating benchmarks test energy transfer from the visible sector back into the dark thermal bath in the false vacuum.
For positive heat capacity, heat transfer when $T_\gamma>T_{\rm false}$ opposes cooling in Eq.~\eqref{reh:phasewise-energy} and can suppress further nucleation.
The benchmarks $E_1$ and $E_4$ use $10g_{e,0}$ and $10^4g_{e,0}$, respectively, while $M_4$ uses $10^4g_{\mu,0}$; the other lepton coupling remains at its B1 value.
We include both lepton channels for every benchmark in this comparison, including B1.
The static muon corrections are evaluated with vanishing renormalized vacuum $\phi$--$H_D$ mixing at $Q=\mu_{\rm GW}$.
Their small size on the tested bubble profiles does not bound independent ultraviolet matching or prefactor corrections.
Figure~\ref{fig:visible-portal-controls} shows the phase-conversion and temperature histories, including the intermediate coupling variations.

Inside the bubbles, the heat transferred by scalar decays and inverse decays depends on both the decay widths and the thermal abundances in Eq.~\eqref{reh:transfer-moment}.
Even when $H_D$ and $A_S$ decay rapidly to $\mu^+\mu^-$, their thermal abundances fall more steeply as the baths cool than that of the lighter $\sigma$.
Outside the bubbles, the same coupling that sets the $\sigma\to e^+e^-$ width also allows electron--positron annihilation and elastic scattering to heat the symmetric phase at temperatures where thermal muons are rare [Eq.~\eqref{reh:false-transfer}].
In $E_4$, this heating suppresses further nucleation when few bubbles have formed, and part of the volume remains in the false vacuum at the end of the evolution.
The transition completes for the weaker electron coupling in $E_1$.

In $M_4$, once the visible bath reheats, muon pairs annihilating into $\phi H_D$ or $\phi A_S$ supply most of the heat transferred to the symmetric phase.
By then enough bubbles have formed to complete conversion as they expand, even though heating suppresses further nucleation.
The larger bubble separation in Hubble units, $H_*R_*$, shifts the GW peak to lower frequency and higher amplitude relative to B1.
Because the charge is fixed and the entropy changes only modestly, the relic abundance remains close to its B1 value at the original $r_X$.

\subsection{Numerical checks and limitations of the nucleation rate}
\label{sec:rate-diagnostics}
For B1, the first-law and entropy residuals and the drift in conserved charge are below $10^{-9}$.
An independent integration reproduces the number of bubbles to relative accuracy $10^{-6}$, while refining the evolution changes the calculated spectra by less than $2\times10^{-5}$ for B1 and the other benchmarks with the same AD parameters.
These numerical checks do not quantify the theoretical uncertainty of the nucleation rate.

For B1, $S$ is not much heavier than $\phi$ in the false vacuum, and the mass ratios used in the high-temperature expansion are not uniformly small across the bubble.
We repeated the B1 evolution at fixed $r_X$ with the local finite-$m_\phi$ correction to the mixed scalar diagram evaluated on the original B1 bounce profiles, subtracting its leading overlap with the radial fluctuation determinant.
The transition still completes, with only small shifts in the completion time and the mean bubble separation.
The rate omits hard-mode potential terms beyond quartic order, additional real-scalar soft terms, field-dependent kinetic corrections and plasma dissipation.
Its perturbative treatment differs from that of the homogeneous phase thermodynamics, and the omitted soft and derivative terms have not been controlled within a common power counting.
The effects of varying the matching scales and of the remaining omitted terms on the thermal histories are unquantified.
The tabulated variations also leave the effects of finite-density rate corrections entangled with those of the AD charge, energy and thermal evolution.
These selected benchmarks illustrate how the predictions depend on the parameters.
A systematic parameter scan and a theoretical uncertainty estimate lie beyond the scope of this comparison.

\section{Compatibility with Big Bang Nucleosynthesis}
\label{sec:phenomenological-implications}
\label{sec:bbn-delayed-heating}
Bubble walls reach different regions at different times, so a first-order transition can produce spatial variations in the heat deposited in the plasma~\cite{Elahi:2026ldn}.
To test the effect of nonlinear transport on nucleosynthesis, we start with a cosine perturbation of the initial temperatures and evolve it on the respective expansion histories of B1, $C_v$, $C_y$, $C_{\phi S}$, $C_S$, $C_Q$, $E_1$ and $M_4$.
The initial baryon density is uniform and the fluids are initially at rest.
We choose the inverse comoving mean bubble separation as the wavenumber.
We set the amplitude of the perturbation in the local conversion time equal to the standard deviation of that time in the bubble nucleation model.
Appendix~\ref{app:bbn-details} gives the initial temperature perturbation in Eq.~\eqref{bbn:cosine-initial} and the fluid and neutrino transport equations that govern its evolution.

Entropy production during visible-sector reheating dilutes any pre-existing baryon asymmetry.
On the homogeneous background, define the baryon yield $Y_B=(n_B-n_{\bar B})/s_{\rm vis}$ and the comoving visible-sector entropy $S_{\rm vis}=a^3s_{\rm vis}$, including the neutrino contribution.
If baryogenesis precedes reheating and the net baryon number in a comoving volume is conserved thereafter, the required yield before entropy injection exceeds the yield after it by the dilution factor $\Delta_{\rm vis}$,
\begin{equation}
Y_B^{\rm pre}=\Delta_{\rm vis}\,Y_B^{\rm post},
\qquad
\Delta_{\rm vis}=\frac{S_{\rm vis}^{\rm post}}{S_{\rm vis}^{\rm pre}}>1 .
\label{bbn:baryon-dilution}
\end{equation}
We normalize the BBN histories to the observed late-time baryon density $\omega_b=\Omega_b h^2$, taking the baryon asymmetry as an input.
For B1, the dilution factor $\Delta_{\rm vis}=1.515\times10^6$ in Table~\ref{tab:benchmark-summary} implies that a late baryon yield $Y_B^{\rm post}\simeq8.7\times10^{-11}$ requires $Y_B^{\rm pre}\simeq1.3\times10^{-4}$ before the phase transition.
Producing this initial baryon yield is an additional requirement on the cosmological history; the AD evolution considered here generates the dark asymmetry, while the visible baryon asymmetry is specified independently.

We follow fluid elements through the nonlinear flow, each carrying a fixed fraction of the total baryon number.
Along each trajectory, we supply the local photon and electron-neutrino temperatures, baryon density and proper time to LINX~\cite{Giovanetti:2024zce}.
We run this network separately with the PArthENoPE rate fits~\cite{Gariazzo:2021iiu,Pisanti:2020efz} and the PRIMAT 2023 rate tables~\cite{Pitrou:2018cgg}.
Equation~\eqref{bbn:mixed-yields} gives D/H from the total numbers of deuterons and free protons, and the helium mass fraction $Y_{\rm P}$ from the total number of \ce{^4He} nuclei per baryon.

We compare the abundance vector $\mathbf y=(\mathrm{D/H},Y_{\rm P})$ with the observed primordial abundances~\cite{ParticleDataGroup:2026mpi} in Table~\ref{tab:benchmark-history-bbn}.
For each rate prescription, Eq.~\eqref{bbn:joint-comparison} combines the independent observational errors with the abundance covariance from nuclear rates, the neutron lifetime and the mean baryon density at fixed transport histories.
The two prescriptions are implemented in the same LINX network and give separate theoretical predictions.
With two measured abundances and no parameters fitted to them, the adopted joint $95\%$ criterion is $\chi^2_{\rm BBN}<5.991$.

\begin{table}[!htbp]
\centering
\small
\begin{tabular}{lrrr}
\toprule
Prediction or observation & $10^5(\mathrm{D/H})$ & $Y_{\rm P}$ & $\chi^2_{\rm BBN}$ \\
\midrule
B1 (PArthENoPE) & 2.562 & 0.2446 & 1.116 \\
B1 (PRIMAT) & 2.492 & 0.2447 & 0.1203 \\
\addlinespace
$C_v$ (PArthENoPE) & 2.565 & 0.2444 & 1.256 \\
$C_v$ (PRIMAT) & 2.495 & 0.2446 & 0.09323 \\
\addlinespace
$C_y$ (PArthENoPE) & 2.599 & 0.2430 & 3.478 \\
$C_y$ (PRIMAT) & 2.528 & 0.2431 & 0.5633 \\
\addlinespace
$C_{\phi S}$ (PArthENoPE) & 2.549 & 0.2451 & 0.6470 \\
$C_{\phi S}$ (PRIMAT) & 2.480 & 0.2452 & 0.3549 \\
\addlinespace
$C_S$ (PArthENoPE) & 2.569 & 0.2443 & 1.447 \\
$C_S$ (PRIMAT) & 2.499 & 0.2444 & 0.07685 \\
\addlinespace
$C_Q$ (PArthENoPE) & 2.562 & 0.2446 & 1.117 \\
$C_Q$ (PRIMAT) & 2.492 & 0.2447 & 0.1201 \\
\addlinespace
$E_1$ (PArthENoPE) & 2.589 & 0.2434 & 2.742 \\
$E_1$ (PRIMAT) & 2.519 & 0.2435 & 0.2919 \\
\addlinespace
$M_4$ (PArthENoPE) & 2.614 & 0.2425 & 4.875 \\
$M_4$ (PRIMAT) & 2.544 & 0.2426 & 1.189 \\
\midrule
Observation & $2.508\pm0.029$ & $0.245\pm0.003$ & \\
\bottomrule
\end{tabular}
\caption{Primordial deuterium and \ce{^4He} abundances obtained by evolving one-dimensional cosine perturbations of the initial temperatures in benchmarks in \ref{tab:benchmark-summary} except E4.
Each prediction uses the indicated LINX rate prescription, and the final row gives the observed abundances with their $1\sigma$ uncertainties.}
\label{tab:benchmark-history-bbn}
\end{table}

All sixteen predictions in Table~\ref{tab:benchmark-history-bbn} satisfy the joint abundance criterion, with $0.07685\leq\chi^2_{\rm BBN}\leq4.875$.
The PArthENoPE predictions for D/H lie above the observed central value for all eight benchmarks.
The PRIMAT predictions lie below it for B1, $C_v$, $C_{\phi S}$, $C_S$ and $C_Q$ and above it for $C_y$, $E_1$ and $M_4$, while the helium predictions lie within the observational $1\sigma$ interval in every case.
Changing the rate prescription shifts D/H by approximately $7.0\times10^{-7}$ in each case, and we use the predictions from each rate prescription in separate joint tests.

The abundance predictions respond much less to these benchmark variations than the asymmetric relic density does.
Relative to B1 at fixed rate prescription, $C_S$ increases D/H by about $0.27\%$, while $C_Q$ gives the same deuterium and helium abundances to the precision shown in the table.
The scalar-portal variation $C_{\phi S}$ decreases D/H by about $0.50\%$.
The Yukawa variation $C_y$ increases D/H by $1.43$--$1.45\%$ across the two prescriptions.
The muon-portal benchmark $M_4$ gives the largest joint statistic, using PArthENoPE.
The corresponding relic abundances in Table~\ref{tab:benchmark-summary} range from $\Omega_\chi h^2=0.01385$ to $0.4041$, compared with $0.12$ for B1.
Thus the sensitivity of the relic abundance to charge dilution does not translate into a comparable change in the mixed light-element abundances for the temperature histories obtained from these initial cosine perturbations. 
Their covariance describes the propagated nuclear-rate, neutron-lifetime and baryon-density inputs; uncertainty in the transport model and its closure is not included.
Within this scope, nonlinear evolution of the specified perturbations remains compatible with the observed abundances.
Extending the result to the spatially averaged abundances of the phase transition requires the correlated heating field generated by stochastic bubble nucleation and its subsequent transport.
\FloatBarrier

\section{Conclusion}
\label{sec:conclusions}
We studied a supercooled first-order phase transition in a classically conformal dark sector with asymmetric dark matter.
In this model, radiative symmetry breaking generates masses and an Affleck--Dine condensate generates a fermion asymmetry and heats the dark bath before the transition.
The coupled thermal evolution sets the mean bubble separation and the final electromagnetic entropy.
For a given conserved dark charge, this entropy fixes the asymmetric relic abundance and, together with the bubble separation, determines the GW spectrum.

With $r_X$ fixed by the relic-density calibration, the B1 transition completes for the nucleation rate used in this work.
Under the assumed dominance of bubble collisions, the GW spectrum peaks at nanohertz frequencies.

For the initial cosine temperature perturbations evolved in B1, $C_v$, $C_y$, $C_{\phi S}$, $C_S$, $C_Q$, $E_1$ and $M_4$, both nuclear-rate prescriptions satisfy the joint $95\%$ deuterium and helium criterion with the input covariance evaluated at fixed transport histories.
The transport and closure limitations described above also apply to this conditional compatibility, and extending it to the phase transition requires evolving the correlated heating field generated by stochastic bubble nucleation.
The late baryon density also imposes a requirement on the preceding history: baryogenesis before reheating must supply an asymmetry large enough to survive the visible-sector entropy dilution.

The benchmarks with varied dark-sector parameters show that changes in the transition time strongly affect charge dilution, while the GW peak is less sensitive.
Nearby variations of $r_X$ mainly affect the relic abundance.
Also, energy transfer back to the dark thermal bath in the false vacuum makes nucleation and transition completion sensitive to the reheating couplings.

\acknowledgments
We thank Jorinde van de Vis for helpful discussions on the nucleation calculation in the three-dimensional effective theory, including scalar fluctuations and the choice of renormalization scales.
We also thank Philipp Schicho for providing independent DRalgo calculations of the matching coefficients and hard pressure in our model at zero and nonzero fermion chemical potential.
W.C.\ and K.-Y.C.\ acknowledge financial support from a National Research Foundation (NRF) grant funded by the Korean government (MEST), NRF-2022R1A2C1005050.
W.C.\ acknowledges support from the CERN--Korea Theoretical Physics Collaboration and Developing Young High-Energy Theorists fellowship program (NRF-2012K1A3A2A0105178151).
\par

\appendix
\section{Effective potential and thermodynamics}
\label{app:thermal-inputs}
\subsection{Matching coefficients at finite chemical potential}
\label{app:density-matching}
We use the finite-density sum-integrals of Refs.~\cite{Vuorinen:2003fs,Gynther:2003za}.
For the fermion vector chemical potential $\mu_\chi$, we define $\eta_\chi=\mu_\chi/T_D$ and $\Delta F\equiv F(\eta_\chi)-F(0)$ at fixed four-dimensional $\overline{\rm MS}$ couplings and matching scales.
With $\psi$ the digamma function, the one-loop hard field-normalization corrections $\Delta z_i$ and canonically normalized quartic corrections are
\begin{equation}
\begin{split}
\Delta z_\phi&=2y_\chi^2\Delta\widetilde I_2,\qquad \Delta\widetilde I_2=-\frac{\operatorname{Re}\psi(\frac12+\frac{i\eta_\chi}{2\pi})-\psi(\frac12)}{8\pi^2},\\
\frac{\Delta\lambda_{\phi,3}}{\TD}&=4y_\chi^2(y_\chi^2-\lambda_\phi)\Delta\widetilde I_2,\qquad \frac{\Delta\lambda_{\phi S,3}}{\TD}=-\lambda_{\phi S}\Delta z_\phi.
\end{split}
\label{mod:density-hard-differences}
\end{equation}
The corresponding complex-scalar corrections satisfy $\Delta z_S=\Delta\lambda_{S,3}=0$.

For the two-loop masses, let $\nu=\eta_\chi/(2\pi)$, $q(u)=\operatorname{Re}\psi(\tfrac12+iu)$, $K=(16\pi^2)^{-1}$ and $L=\ln[\bar\mu_4/(4\pi T_D)]$.
With $\gamma_E$ the Euler--Mascheroni constant, the two functions entering the masses are
\begin{equation}
\begin{split}
D_\mu&=4K\int_0^\nu\dd u\,u[1-\gamma_E-q(u)],\\
Y_\mu&=\frac{\Delta\widetilde I_2}{6}-8D_\mu+4K\nu^2[q(\nu)-4\gamma_E-5L].
\end{split}
\label{mod:density-reduced-functions}
\end{equation}
The canonically normalized two-loop mass corrections, with $b_B=1/12$, are
\begin{equation}
\begin{split}
\frac{\Delta m_{\phi,3}^{2,(2)}}{T_D^2}&=y_\chi^4Y_\mu+6\lambda_\phi y_\chi^2(D_\mu-b_B\Delta\widetilde I_2)-2\lambda_{\phi S}y_\chi^2b_B\Delta\widetilde I_2,\\
\frac{\Delta m_{S,3}^{2,(2)}}{T_D^2}&=\lambda_{\phi S}y_\chi^2D_\mu.
\end{split}
\label{mod:density-evaluated-masses}
\end{equation}
Both real components of $S$ receive the same correction.
These mass terms supplement the zero-density matching and the leading density correction in Eq.~\eqref{mod:finite-density-mass}.

\subsection{Effective potential at finite chemical potential}
\label{app:phase-eos}
The bulk grand-potential densities at finite temperature and chemical potential give the pressure, charge density and energy density of each phase through Eq.~\eqref{pt:phase-thermodynamics}.
The false-vacuum-subtracted bounce potential alone does not determine these quantities; bubble translation and growth factors in the nucleation rate do not contribute to bulk pressure.

The symmetric-phase grand potential combines the hard unit-operator term $V_{{\rm eff},{\rm hard}}$ through two loops~\cite{Braaten:1995cm} and homogeneous static terms, with $a_{\rm eff}=a_\phi-\ell_{\phi S}\sqrt{a_S}/(4\pi)$ and all matched coefficients evaluated at $T_{\rm false}$:
\begin{equation}
\begin{split}
V_{{\rm eff},\mathrm{false}}={}&\Delta V_0+V_{{\rm eff},{\rm hard}}+T_{\mathrm{false}}^4\left[-\frac{2a_S^{3/2}+a_{\rm eff}^{3/2}}{12\pi}+\frac{\ell_Sa_S}{8\pi^2}\right],\\
\frac{V_{{\rm eff},{\rm hard}}}{T_{\mathrm{false}}^4}={}&-\frac{13\pi^2}{180}-\frac{\eta_{\chi,\mathrm{false}}^2}{6}-\frac{\eta_{\chi,\mathrm{false}}^4}{12\pi^2}+\frac{\lambda_\phi}{192}+\frac{\lambda_S}{72}+\frac{\lambda_{\phi S}}{288}\\
&+y_\chi^2\left(\frac{5}{576}+\frac{\eta_{\chi,\mathrm{false}}^2}{32\pi^2}+\frac{\eta_{\chi,\mathrm{false}}^4}{64\pi^4}\right).
\end{split}
\label{pt:false-grand-potential}
\end{equation}
The local complex-scalar and mixed scalar double-bubble terms are each counted once; thermodynamic derivatives include the matched coefficients' running and density dependence.

For $z=m^2/T^2$ and $\eta_\chi=\mu_\chi/T$, the thermal functions in the broken-phase potential of Eq.~\eqref{pt:phase-grand-potentials} are
\begin{equation}
\begin{split}
J_B(z)={}&\int_0^\infty\dd x\,x^2\ln\!\left(1-e^{-\sqrt{x^2+z}}\right),\\
J_F(z,\eta_\chi)={}&\frac12\int_0^\infty\dd x\,x^2\left[\ln\!\left(1+e^{-\sqrt{x^2+z}+\eta_\chi}\right)+\ln\!\left(1+e^{-\sqrt{x^2+z}-\eta_\chi}\right)\right].
\end{split}
\label{mod:thermal-integrals}
\end{equation}
We use the positive-curvature branch near $v$ satisfying $\partial_\phi V_{\rm eff,true}=0$, including its field variation in second thermodynamic derivatives.
We omit separate daisy terms and interaction corrections among massive broken-phase particles.

With the common vacuum normalization $U(v)=0$ and $U(0)=\Delta V_0$, the pressure difference $\Delta V$ and released energy density $\Delta e$ at common $T_D$ and $\eta_\chi$ are
\begin{equation}
\Delta V=V_{{\rm eff},\mathrm{false}}-V_{{\rm eff},\mathrm{true}},\qquad \Delta e=\Delta V-T_D\left.\frac{\partial\Delta V}{\partial T_D}\right|_{\eta_\chi}.
\label{pt:release}
\end{equation}
Both quantities approach $\Delta V_0$ in the strongly supercooled limit.
Here we evaluate both phases at a common temperature, whereas in Section~\ref{sec:reh:reheating} we evolve their temperatures separately.

\section{Big Bang nucleosynthesis}
\label{app:bbn-details}
\subsection{Temperature perturbations and nucleosynthesis}

We take the expansion rate $H(N)$ from the homogeneous reheating history of each benchmark, including continuous thermal neutrino energy exchange~\cite{Escudero:2018mvt,EscuderoAbenza:2020cmq}, where $N=\ln(a/a_p)$.
Our background and spatial transport calculations use the neutrino energy-transfer routines and the finite-temperature QED pressure table from \texttt{nudec\_BSM\_v2}~\cite{Escudero:2025kej}.
Let $N_c$ be the value of $N$ at which a bubble wall first reaches a given point and $\sigma_{N_c}$ its standard deviation, obtained from the survival probability $P_{\rm false}=e^{-I}$ in Eq.~\eqref{pt:percolation}.
At the initial time $N_0$, on the final cooling branch with background photon temperature $\bar T_\gamma=10\MeV$, we set the temperature of each component $i$ to
\begin{equation}
T_i(N_0,x)=\bar T_i(N_0)\bigl[1+s_i\sigma_{N_c}\cos(kx)\bigr],
\qquad s_i=\left.\frac{\partial\ln T_i}{\partial N_c}\right|_0 .
\label{bbn:cosine-initial}
\end{equation}
Here $x$ is a comoving coordinate and $k=a_p/R_*$ is the inverse comoving bubble separation, with $R_*$ defined at percolation by Eq.~\eqref{pt:bubble-density}.
We evolve one period $2\pi/k$ with periodic boundary conditions.
We assign both neutrino flavour groups the initial photon slope as a matching assumption, since their separate conversion-time responses have not been measured.
The cosine displacement in $N_c$ has peak amplitude $\sigma_{N_c}$ and spatial root-mean-square amplitude $\sigma_{N_c}/\sqrt{2}$.
Initially the bulk velocities vanish, the neutrino intensities are isotropic and the baryon density, dark charge density and minority-particle density are uniform.
We normalize the comoving baryon number to the homogeneous background and use this normalization when comparing the abundances obtained by evolving this initial cosine perturbation with observations.

We evolve the electromagnetic and dark fluids on the unperturbed expanding background, using $\theta=kx$ and $\kappa=k/(aH)$.
For each fluid $i\in\{\mathrm{EM},D\}$, let $v_i$, $\gamma_i=(1-v_i^2)^{-1/2}$ and $h_i=\rho_i+p_i$ denote its velocity, Lorentz factor and rest-frame enthalpy density, respectively.
In the cosmological frame the energy density, momentum density and longitudinal stress are $E_i=h_i\gamma_i^2-p_i$, $F_i=h_i\gamma_i^2v_i$ and $\Pi_i=F_iv_i+p_i$.
With energy and momentum sources $Q_i^0$ and $Q_i^x$ in this frame, the fluid equations are
\begin{equation}
\begin{split}
\partial_N E_i+\kappa\,\partial_\theta F_i
 &=-3E_i-\Pi_i-2p_i+Q_i^0/H,\\
\partial_N F_i+\kappa\,\partial_\theta\Pi_i
 &=-4F_i+Q_i^x/H .
\end{split}
\label{bbn:fluid-transport}
\end{equation}
The portal transfers four-momentum to the electromagnetic fluid at the rate $Q_{\rm portal}^\mu=q(T_D)u_D^\mu-q(T_\gamma)u_{\rm EM}^\mu$, with the opposite source for the dark fluid.
Here $u_i^\mu$ is the fluid four-velocity and $q(T)=\sum_s m_s\Gamma_{s\to{\rm SM}}n_s^{\rm MB}(T)$ is the decay contribution in Eq.~\eqref{reh:transfer-moment}.

For each flavour group $f\in\{e,\mu\tau\}$, we transport the combined neutrino and antineutrino intensity $I_f(\mu)$, integrated over frequency and measured per unit direction cosine $\mu$ in the cosmological frame.
The energy density in the electromagnetic rest frame is $J_f=\gamma_{\rm EM}^2\int_{-1}^1(1-v_{\rm EM}\mu)^2 I_f(\mu)\,\dd\mu$.
We define the effective neutrino temperature by $J_f=d_f(7\pi^2/120)T_{\nu_f}^4$, with $d_e=1$ and $d_{\mu\tau}=2$.
Writing $\mathcal D_\mu=[\gamma_{\rm EM}(1-v_{\rm EM}\mu)]^{-1}$, we use the frequency-integrated collision model
\begin{equation}
\begin{split}
(\partial_N+\kappa\mu\,\partial_\theta)(a^4I_f)&=\frac{a^4}{H}\mathcal C_f(\mu),\\
\mathcal C_f(\mu)&=\frac{\Gamma_\nu J_f+w_f}{2}\mathcal D_\mu^3
 -\frac{\Gamma_\nu}{\mathcal D_\mu}I_f(\mu).
\end{split}
\label{bbn:neutrino-transport}
\end{equation}
Here $w_f$ is the net energy gained by the flavour group per unit proper volume and time from the thermal weak rates, evaluated at the local $T_\gamma$, $T_{\nu_e}$ and $T_{\nu_{\mu\tau}}$.
The angular relaxation rate is $\Gamma_\nu=G_F^2T_\gamma^5\,n_{e^\pm}^{\rm eq}(T_\gamma)/[3\zeta(3)T_\gamma^3/\pi^2]$, where $G_F$ is the Fermi constant and $n_{e^\pm}^{\rm eq}$ counts equilibrium electrons and positrons at zero chemical potential.
We subtract the energy and momentum moments $\sum_f\int_{-1}^1(1,\mu)\mathcal C_f\,\dd\mu$ from the electromagnetic sources in Eq.~\eqref{bbn:fluid-transport}.

At a background photon temperature of $1.5\MeV$, we switch off dark-sector evolution and portal exchange and include photon momentum diffusion.
In this stage the electromagnetic momentum equation acquires the term $A_\gamma\partial_\theta^2 F_{\rm EM}$, where
\begin{equation}
A_\gamma=\frac{16}{45}\frac{4\bar\rho_\gamma}{3\bar h_{\rm EM}}\frac{(k/a)^2}{H\Gamma_\gamma}.
\label{bbn:photon-diffusion}
\end{equation}
We evaluate the photon density $\bar\rho_\gamma$, electromagnetic enthalpy density $\bar h_{\rm EM}$ and photon scattering rate $\Gamma_\gamma$ on the homogeneous history, using a Klein--Nishina correction to Thomson scattering on pairs and residual electrons.
At $80\,\mathrm{keV}$ we end weak energy exchange and subsequently hold $a^4I_f(\mu)$ at its spatial mean at that epoch for each angle.

The baryon current follows the electromagnetic fluid, with microscopic nucleon diffusion neglected.
Let $\mathcal B=\gamma_{\rm EM}n_b/\bar n_b$, where $\bar n_b\propto a^{-3}$ is the homogeneous proper baryon density.
We have $\partial_N\mathcal B+\kappa\,\partial_\theta(\mathcal B v_{\rm EM})=0$, with $\mathcal B=1$ initially.
Exact reflection symmetry gives zero flux at $\theta=0$; for trajectories with recorded boundary currents we include their signed cumulative contribution $\mathcal F_0(N)=\int_{N_0}^{N}\kappa(N')(\mathcal B v_{\rm EM})(N',0)\,\dd N'$.
We locate the fluid element with initial coordinate $\theta_{0j}$ by integrating the conserved baryon density on the spatial grid.
Its position $\theta_j(N)$, local proper baryon density $n_{b,j}$ and proper time $\tau_j$ satisfy
\begin{equation}
\begin{split}
\int_0^{\theta_j(N)}\mathcal B(N,\vartheta)\,\dd\vartheta&=\theta_{0j}+\mathcal F_0(N),\\
n_{b,j}=\bar n_b\,\frac{\mathcal B_j}{\gamma_j},\qquad
\frac{\dd\tau_j}{\dd N}&=\frac{1}{H\gamma_j}.
\end{split}
\label{bbn:baryon-tracer}
\end{equation}
where $\mathcal B_j$ and $\gamma_j$ are evaluated at $\theta_j(N)$, with $\gamma_j$ the electromagnetic fluid Lorentz factor there.
At the position of each fluid element, we recover $T_{\gamma,j}$ from the electromagnetic equation of state and $T_{\nu_e,j}$ from $J_e$.
Along each trajectory, LINX integrates $\dd Y_{A,j}/\dd\tau_j=\mathcal R_A(T_{\gamma,j},T_{\nu_e,j},n_{b,j},\boldsymbol Y_j)$, where $\mathcal R_A$ is the net reaction rate for the abundance $Y_{A,j}=n_{A,j}/n_{b,j}$.
For neutron--proton interconversion, LINX implements the PRIMAT weak-rate formalism~\cite{Pitrou:2018cgg} and uses finite-temperature radiative and bremsstrahlung corrections tabulated with PRyMordial~\cite{Burns:2023sgx}.
During deuterium burning, the PRIMAT tables use the $d(p,\gamma){}^3\mathrm{He}$ rate of Ref.~\cite{Moscoso:2021xog} and the $d(d,n){}^3\mathrm{He}$ and $d(d,p)t$ rates of Ref.~\cite{InestaGomez:2017eya}.

We start neutron--proton interconversion at a background photon temperature of $8\MeV$ and keep the evolved neutron and proton abundances when initializing the heavier nuclei in local nuclear statistical equilibrium at $0.8617\MeV$.
We then evolve the eight-species network to a background photon temperature of $1.020\,\mathrm{keV}$.
We combine the abundances from the individual trajectories using weights $w_j$ equal to the fractions of the total baryon number carried by the corresponding fluid elements.
With $\sum_j w_j=1$, the predicted abundances are
\begin{equation}
\mathrm{D/H}=\frac{\sum_j w_jY_{d,j}}{\sum_j w_jY_{p,j}},\qquad
Y_{\rm P}=4\sum_j w_jY_{\alpha,j},
\label{bbn:mixed-yields}
\end{equation}
where $d$, $p$ and $\alpha$ denote deuterons, free protons and \ce{^4He} nuclei.
The calculation uses 128 spatial cells, 48 neutrino directions and eight baryon trajectories.

\subsection{Comparison with observations}

For each rate prescription, let $\mathbf y=(\mathrm{D/H},Y_{\rm P})$ be the prediction from Eq.~\eqref{bbn:mixed-yields}.
We use twelve independent Gaussian coordinates for the LINX lognormal reaction-rate variations~\cite{Giovanetti:2024zce} and two additional independent coordinates for the neutron lifetime $\tau_n=(878.4\pm0.5)\,\mathrm{s}$~\cite{ParticleDataGroup:2026mpi} and the physical baryon density $\omega_b=\Omega_bh^2=0.02237\pm0.00015$~\cite{Planck:2018vyg}.
Each variation is shared by all trajectories at fixed transport histories and fixed weights.
Varying $\omega_b$ rescales the proper baryon density by the same factor in the nuclear initial conditions and throughout the reaction network.
For each input $\theta_j$ with standard deviation $\sigma_j$, the abundance response $\Delta_j\mathbf y$ contributes to the first-order input covariance $C_{\rm input}$.
With the observations $\mathbf y_{\rm obs}$ and their covariance $C_{\rm obs}$ from Section~\ref{sec:bbn-delayed-heating}, the joint statistic is given by
\begin{equation}
\begin{split}
\Delta_j\mathbf y={}&\frac{\mathbf y(\theta_j+\sigma_j)-\mathbf y(\theta_j-\sigma_j)}{2},\qquad C_{\rm input}=\sum_j\Delta_j\mathbf y\,\Delta_j\mathbf y^\top,\\
\chi_{\rm BBN}^2={}&(\mathbf y-\mathbf y_{\rm obs})^\top(C_{\rm obs}+C_{\rm input})^{-1}(\mathbf y-\mathbf y_{\rm obs}) .
\end{split}
\label{bbn:joint-comparison}
\end{equation}
For $E_1$, both prescriptions pass the tested abundance refinements, while the late neutrino perturbations fail angular convergence and the photon acoustic phase remains sensitive to spatial resolution.
These field and closure limitations remain outside $C_{\rm input}$.

\providecommand{\href}[2]{#2}\begingroup\raggedright\endgroup

\end{document}